\documentclass{raa}

\usepackage{graphicx,times}             
\usepackage{longtable}
\usepackage{natbib}
\usepackage{amssymb,amsmath}
\bibpunct{(}{)}{;}{a}{}{,}

\usepackage[pagebackref=true]{hyperref}
\begin{document}
\defcitealias{ngc5904}{Paper I}
\defcitealias{ngc6205}{Paper II}
\defcitealias{ngc288}{Paper III}
\defcitealias{ngc6362}{Paper IV}
\defcitealias{ngc6397}{Paper V}
\defcitealias{ngc5024}{Paper VI}
\defcitealias{paper7}{Paper VII}
\defcitealias{nardiello2018}{NLP18}
\defcitealias{stetson2019}{SPZ19}
\defcitealias{vasiliev2021}{VB21}
\defcitealias{baumgardt2021}{BV21}
\defcitealias{sfd98}{SFD98}
\defcitealias{gms2025}{GMS25}
\defcitealias{green2019}{GSZ19}
\defcitealias{aguado2025}{CARMA}
\defcitealias{tailo2020}{TML20}

\title{Isochrone Fitting of Galactic Globular Clusters -- VIII. Homogeneous estimates of parameters for 27 clusters}

   \volnopage{Vol.0 (20xx) No.0, 000--000}      
   \setcounter{page}{1}          

\author{G.~A.~Gontcharov
\inst{1,2}
\and O.~S.~Ryutina
\inst{2,3}
\and C.~J.~Bonatto
\inst{4}
\and S.~S.~Savchenko
\inst{3}
\and S.~V.~Troitsky
\inst{2,5}
}

\institute{Central (Pulkovo) Astronomical Observatory, Russian Academy of Sciences, Pulkovskoye chaussee 65/1, St. Petersburg 196140, Russia; {\it georgegontcharov@yahoo.com}\\
\and
Institute for Nuclear Research, Russian Academy of Sciences, Moscow, 117312 Russia\\
\and
Saint Petersburg State University, 7/9 Universitetskaya nab., St. Petersburg, 199034 Russia\\
\and
Departamento de Astronomia, Instituto de F\'isica, UFRGS, Av. Bento Gon\c{c}alves, 9500, Porto Alegre, RS, Brazil\\
\and
Faculty of Physics, Moscow State University, Moscow, 119991 Russia\\
\vs\no
   {\small Received 20xx month day; accepted 20xx month day}}

\abstract{We use the Stetson, {\it Hubble Space Telescope (HST)}, and {\it Gaia} Data Release 3 data sets for each of 22 Galactic globular clusters to select their members and fit their colour--magnitude diagrams with isochrones from the Dartmouth Stellar Evolution Database and a Bag of Stellar Tracks and Isochrones for $\alpha$--enrichment $[\alpha/$Fe$]=+0.4$ and a helium mass fraction $Y$ adopted for a combination of stellar generations. As a result, we estimate the metallicity $[$Fe$/$H$]$, age, distance from the Sun, and reddening $E(B-V)$ for these clusters. Special attention is paid to identify variable stars among the cluster members, since they affect the derived parameters and statistics.
We combine these results with our earlier estimates for five other clusters into a homogeneous set of parameters for 27 clusters and investigate relationships among these parameters and their statistical properties. In particular, we count the giants on the red (RGB), horizontal (HB), and asymptotic branches using the clean data sets cross-identified to cover the entire cluster fields and count each giant only ones. This allows us to calculate the $R$-parameter, the ratio between the numbers of the HB and RGB, with unprecedented accuracy and to examine its relations with other parameters. These relations are much stronger for accreted clusters rejecting a constant $R$ value for them, possibly because of their heterogeneous origins and environments. 
For all in-situ clusters, a constant value of $R=1.31^{+0.06}_{-0.05}$ is consistent with the data. 
According to previous calculations, this value may imply anomalous energy losses in HB stars corresponding to an axion--photon coupling of $g_{a\gamma}=(0.66^{+0.11}_{-0.13})\times10^{-10}\,\mathrm{GeV}^{-1}$. However, the observed dependencies of $R$ on cluster parameters were not accounted for in previous studies. Any future study of the $R$ parameter aimed at constraining exotic particle-physics models will therefore require not only improved statistics, but also a theoretical understanding of these dependencies and their incorporation into the analysis.
By comparing $R$-parameter estimates from {\it HST} and {\it Gaia} in cluster centres and peripheries, respectively, we suggest that the peripheral HB population is depleted by about half when the cluster crosses the Galactic disk, then the HB recovers over 60--80 Myr.
}

\authorrunning{G. A. Gontcharov et al.}            
\titlerunning{Isochrone fitting of Galactic globular clusters (VIII)}  
\maketitle

\section{Introduction}
\label{intro}

Galactic globular clusters are natural laboratories for verification of theoretical models of stellar structure, evolution, and dynamics against observations, as well as clarification of the metallicity, age, reddening and distance scales in the Galaxy.
Unfortunately, such comparisons of theory and observations are distorted by systematic errors.

Recently, the data sets obtained with
the {\it Gaia} Data Release 3 (DR3; \citealt{gaiadr3}),
{\it Hubble Space Telescope (HST)} Wide Field Channel of the Advanced Camera for Surveys (ACS) \citep[][hereafter NLP18]{nardiello2018}\footnote{\url{http://groups.dfa.unipd.it/ESPG/treasury.php}},
and
various ground-based telescopes \citep[][hereafter SPZ19]{stetson2019}
\footnote{\url{http://cdsarc.u-strasbg.fr/viz-bin/cat/J/MNRAS/485/3042}, its ongoing updates are presented at
\url{https://www.canfar.net/storage/vault/list/STETSON/homogeneous/Latest_photometry_for_targets_with_at_least_BVI}.}
provide photometry for dozens of clusters.
The dedicated astrometry from {\it HST} \citep{libralato2022}, {\it Gaia} DR3 astrometry, and cross-identification of the \citetalias{stetson2019} data sets with those of {\it Gaia} DR3 allow an accurate selection of cluster members.
Moreover, the {\it HST} data sets cover only a few central arcminutes of the clusters where the {\it Gaia} data sets may have biased photometry, while the {\it Gaia} and \citetalias{stetson2019} data sets cover the remaining outer parts of the cluster fields. Therefore, the combination of the data sets covers the entire cluster fields.
Furthermore, the averaging of results obtained for different filters of the data sets may suppress some systematics.
Thus, in this study, we use these three data sets for each cluster in order to obtain homogeneous estimates of cluster parameters and consider their statistics.

In our previous papers [\citet[][hereafter Paper I]{ngc5904}, \citet[][hereafter Paper II]{ngc6205}, \citet[][hereafter Paper III]{ngc288}, \citet[][hereafter Paper IV]{ngc6362}, \citet[][hereafter Paper V]{ngc6397}, \citet[][hereafter Paper VI]{ngc5024}, and \citet[][hereafter Paper VII)]{paper7},\footnote{All five clusters from \citetalias{ngc5904}--\citetalias{ngc288} are revised in \citetalias{paper7}.}
we have used the {\it HST}, {\it Gaia}, and \citetalias{stetson2019} data sets, in combination with others, for 14 clusters to estimate their metallicity $[$Fe$/$H$]$, age, distance from the Sun $D$, and reddening by fitting their colour--magnitude diagrams (CMDs) with theoretical isochrones derived from stellar evolution models by
Dartmouth Stellar Evolution Database (DSED, \citealt{dotter2007,dotter2008})\footnote{\url{http://stellar.dartmouth.edu/models/}} and
a Bag of Stellar Tracks and Isochrones (BaSTI, \citealt{newbasti,pietrinferni2021})\footnote{\url{http://basti-iac.oa-abruzzo.inaf.it/index.html}}.
The main stages of stellar evolution, namely the main sequence (MS), turn-off (TO), subgiant branch (SGB), red giant branch (RGB), horizontal branch (HB), and asymptotic giant branch (AGB) are successfully reproduced by both the observations and isochrones, except DSED for the HB and AGB.

In this study, we fit the CMDs based on the {\it HST}, {\it Gaia}, and \citetalias{stetson2019} data sets by the BaSTI and DSED isochrones for 13 more clusters, redo the fitting for nine clusters from \citetalias{ngc6362}--\citetalias{ngc5024}, and adopt the results of our fitting of five clusters (NGC\,288, NGC\,362, NGC\,5904, NGC\,6205, and NGC\,6218) from \citetalias{paper7} unchanged. Finally, we consider statistics of various parameters for the sample of 27 clusters.

The reason for repeating the analysis of the nine clusters from \citetalias{ngc6362}--\citetalias{ngc5024} is that we pay more attention to identification of variable stars, in line with our \citetalias{paper7}.
This identification affects our estimates of the cluster parameters.
We identify variable stars using the database of \citet{clement2017}\footnote{\url{https://www.astro.utoronto.ca/~cclement/cat/listngc.html}}, as well as the parameters \verb"Vary" (Welch-Stetson variability index) and \verb"Weight" (weight of the variability index) from the \citetalias{stetson2019} data sets and the parameter \verb"VarFlag=VARIABLE" from the {\it Gaia} data sets.
Involvement of a variable star in or exclusion from our isochrone fitting depends on the type of its variability, as well as on the frequency and number of observations w.r.t. variability period and amplitude, since variable stars observed at random phase may occupy inappropriate positions on CMDs.
For example, we involve a long-period variable star with hundreds of observations compiled in a \citetalias{stetson2019} data set, while we exclude an RR~Lyrae star (with its rather short period) with only a few observations in a {\it Gaia} data set.
Details and results of this identification will be presented in a separate paper.

The coverage of the entire cluster fields by accurate data allows us to count the numbers $N_\mathrm{RGB}$ and $N_\mathrm{HB}$ of the RGB and HB stars, respectively, and calculate their ratio $R\equiv N_\mathrm{HB}/N_\mathrm{RGB}$, also known as $R$-parameter, with unprecedented accuracy.
$R$ can be used as a proxy for the ratio between stellar lifetimes on these stages and, hence, as an estimator of additional energy losses associated with nonstandard particle physics. In particular, these energy losses associated with coupling of photons to hypothetical axion-like particles reduce the HB lifetime, and $R$ was used for constraining this coupling
\citep{raffelt1996,ayala2014,dolan2022,carenza2025,troitsky2025}.

As this is the eighth paper in this series, we refer the reader to our previous papers for the details of our cleaning of the data sets, selection of cluster members, estimation of uncertainties, evaluation of differential reddening, and other stages of obtaining results from our isochrone fitting.

This paper is organized as follows.
Some properties of the clusters under consideration are considered in Sect.~\ref{clusterproperties}.
In Sect.~\ref{newfitting} we present our new isochrone fitting for 22 clusters.
We consider relations of various parameters of the clusters in Sect.~\ref{relations}.
In Sect.~\ref{giants} we present and discuss the count and statistics of the giants.
Our main findings and conclusions are summarized in Sect.~\ref{conclusions}.
All the CMDs with our fitting are shown in Appendix~\ref{addcmds}, the best-fitting parameters are presented in Appendix~\ref{best}, while some more relations between cluster parameters are presented in Appendix~\ref{more}.

\section{Properties of the clusters}
\label{clusterproperties}

Some properties of the clusters under consideration are presented in Table~\ref{properties}.
This information is provided for reference only. More estimates of the parameters from the literature are compared with our estimates in Sect.~\ref{newfitting}.
\footnote{It is worth noting that distance estimates compiled in the commonly used database of clusters by \citet{harris}, 2010 revision (\url{https://www.physics.mcmaster.ca/~harris/mwgc.dat}) are only a part of a comprehensive compilation by \citet{baumgardt2021} of distance estimates obtained by various methods, while reddening $E(B-V)$ estimates in the \citet{harris} database seem to be less reliable w.r.t. those from \citet{sfd98} or \citet{planck} as shown by \citetalias{ngc6362,ngc6397,ngc5024}.
Therefore, the estimates from the \citet{harris} database are not considered in Table~\ref{properties} and Sect.~\ref{newfitting}.}

\begin{table*}
\def\baselinestretch{1}\normalsize\footnotesize
\caption[]{Some properties of the clusters under consideration: \\
Galactic longitude $l$ and latitude $b$ ($^{\circ}$) adopted from \citet{goldsbury2010} or calculated by us as medians for the {\it Gaia} DR3 cluster members,
distance from the Sun $D_\mathrm{BV21}$ (kpc) from \citet{baumgardt2021},
$[$Fe$/$H$]$ from \citet{carretta2009},
age from \citet{dotter2010},
$E(B-V)$ (mag) from \citet{sfd98},
and association from \citet{massari2025}, except NGC\,6397 associated with the low energy subsample following \citet{callingham2022}, [disk, bulge, low energy (low-E), Helmi stream, Gaia-Sausage-Enceladus (GSE), or Sequoia].
}
\label{properties}
\[
\begin{tabular}{lccccccc}
\hline
\noalign{\smallskip}
Cluster & $l$ & $b$ & $D_\mathrm{BV21}$ & $[$Fe$/$H$]$ & Age & $E(B-V)$ & Association \\ 
\hline
\noalign{\smallskip}
NGC\,288  & $151.2815$ & $-89.3804$ &  $8.99\pm0.09$ & $-1.32\pm0.02$ & $12.50\pm0.50$ & 0.01 & GSE \\ 
NGC\,362  & $301.5330$ & $-46.2474$ &  $8.77\pm0.11$ & $-1.30\pm0.04$ & $11.50\pm0.50$ & 0.03 & GSE \\ 
NGC\,1261 & $270.5387$ & $-52.1243$ & $16.40\pm0.19$ & $-1.27\pm0.08$ & $11.50\pm0.50$ & 0.01 & GSE \\ 
NGC\,5024 & $332.9624$ &  $79.7641$ & $18.50\pm0.18$ & $-2.06\pm0.09$ & $13.25\pm0.50$ & 0.02 & Helmi \\ 
NGC\,5053 & $335.6983$ &  $78.9461$ & $17.54\pm0.23$ & $-2.30\pm0.08$ & $13.50\pm0.75$ & 0.02 & Helmi \\ 
NGC\,5272 &  $42.2164$ &  $78.7069$ & $10.18\pm0.08$ & $-1.50\pm0.05$ & $12.50\pm0.50$ & 0.01 & Helmi \\ 
NGC\,5466 &  $42.1499$ &  $73.5923$ & $16.12\pm0.16$ & $-2.31\pm0.09$ & $13.00\pm0.75$ & 0.02 & Sequoia \\ 
NGC\,5897 & $229.3500$ & $-21.0098$ & $12.55\pm0.24$ & $-1.90\pm0.06$ &                & 0.14 & GSE \\ 
NGC\,5904 &   $3.8586$ &  $46.7964$ &  $7.48\pm0.06$ & $-1.33\pm0.02$ & $12.25\pm0.75$ & 0.04 & GSE \\ 
NGC\,6093 & $352.6731$ &  $19.4631$ & $10.34\pm0.12$ & $-1.75\pm0.08$ & $13.50\pm1.00$ & 0.22 & Low-E \\ 
NGC\,6101 & $317.7461$ & $-15.8248$ & $14.45\pm0.19$ & $-1.98\pm0.07$ & $13.00\pm1.00$ & 0.10 & GSE \\ 
NGC\,6171 &   $3.3732$ &  $23.0106$ &  $5.63\pm0.08$ & $-1.03\pm0.02$ & $12.75\pm0.75$ & 0.46 & Bulge \\ 
NGC\,6205 &  $59.0073$ &  $40.9131$ &  $7.42\pm0.08$ & $-1.58\pm0.04$ & $13.00\pm0.50$ & 0.02 & GSE \\ 
NGC\,6218 &  $15.7151$ &  $26.3134$ &  $5.11\pm0.05$ & $-1.33\pm0.02$ & $13.25\pm0.75$ & 0.18 & Disk \\ 
NGC\,6254 &  $15.1370$ &  $23.0761$ &  $5.07\pm0.06$ & $-1.57\pm0.02$ & $13.00\pm1.25$ & 0.29 & Low-E \\ 
NGC\,6341 &  $68.3383$ &  $34.8591$ &  $8.50\pm0.07$ & $-2.35\pm0.05$ & $13.25\pm1.00$ & 0.02 & GSE \\ 
NGC\,6352 & $341.4214$ &  $-7.1661$ &  $5.54\pm0.07$ & $-0.62\pm0.05$ & $13.00\pm0.50$ & 0.37 & Disk \\ 
NGC\,6362 & $325.5545$ & $-17.5697$ &  $7.65\pm0.07$ & $-1.07\pm0.05$ & $12.50\pm0.50$ & 0.07 & Disk \\ 
NGC\,6366 &  $18.4085$ &  $16.0357$ &  $3.44\pm0.05$ & $-0.59\pm0.08$ & $12.00\pm0.75$ & 0.75 & Disk \\ 
NGC\,6397 & $338.1650$ & $-11.9595$ &  $2.48\pm0.02$ & $-1.99\pm0.02$ & $13.50\pm0.50$ & 0.19 & Low-E \\ 
NGC\,6541 & $349.2860$ & $-11.1881$ &  $7.61\pm0.10$ & $-1.82\pm0.08$ & $13.25\pm1.00$ & 0.16 & Low-E \\ 
NGC\,6723 &   $0.0692$ & $-17.2988$ &  $8.27\pm0.10$ & $-1.10\pm0.07$ & $12.75\pm0.50$ & 0.17 & Bulge \\ 
NGC\,6752 & $336.4929$ & $-25.6283$ &  $4.13\pm0.04$ & $-1.55\pm0.01$ & $12.50\pm0.75$ & 0.06 & Disk \\ 
NGC\,6779 &  $62.6593$ &   $8.3365$ & $10.43\pm0.14$ & $-2.00\pm0.09$ & $13.50\pm1.00$ & 0.25 & GSE \\ 
NGC\,6809 &   $8.7925$ & $-23.2715$ &  $5.35\pm0.05$ & $-1.93\pm0.02$ & $13.50\pm1.00$ & 0.14 & Low-E \\ 
NGC\,6838 &  $56.7458$ &  $-4.5643$ &  $4.00\pm0.05$ & $-0.82\pm0.02$ & $12.50\pm0.75$ & 0.33 & Disk \\ 
NGC\,7099 &  $27.1791$ & $-46.8354$ &  $8.46\pm0.09$ & $-2.33\pm0.02$ & $13.25\pm1.00$ & 0.05 & GSE \\ 
\hline
\end{tabular}
\]
\end{table*}
\begin{table}
\def\baselinestretch{1}\normalsize\scriptsize
\caption[]{$Y_\mathrm{mix}$ and $\mu_\mathrm{mix}$ calculated with Equations~(\ref{ymix}) and (\ref{mlmix}), respectively, and adopted for the fitted mix of stellar generations.
The fraction of 1G stars $N_\mathrm{1G}/N_\mathrm{TOT}$ is adopted from \citet{jang2025} or \citet{dondoglio2021},
the primordial $Y_\mathrm{1G}$ is stated by BaSTI for the cluster's metallicity,
the average difference between the generations $\delta Y_\mathrm{2G,1G}$ is adopted from \citet{milone2018},
the mass loss $\mu_\mathrm{1G}$ for 1G (in solar mass), mass-loss difference $\delta \mu_\mathrm{2G,1G}$ between the generations (in solar mass), and mass loss efficiency Reimers parameter $\eta$ are adopted from \citetalias{tailo2020}.
The uncertainties of the input arguments of Equations~(\ref{ymix}) and (\ref{mlmix}) are not shown but can be found in the corresponding publications.
}
\label{mix}
\[
\begin{tabular}{lcccccccc}
\hline
\noalign{\smallskip}
Cluster    & $N_\mathrm{1G}/N_\mathrm{TOT}$ & $Y_\mathrm{1G}$ & $\delta Y_\mathrm{2G,1G}$ & $Y_\mathrm{mix}$ & $\mu_\mathrm{1G}$ & $\delta \mu_\mathrm{2G,1G}$ & $\mu_\mathrm{mix}$ & $\eta$ \\
\hline
\noalign{\smallskip}
NGC\,288   & 0.56 & 0.249 &    0.015 & $0.256\pm0.005$ & $0.213$ & $0.033$ & $0.228\pm0.023$ & $0.516\pm0.021$ \\
NGC\,362   & 0.28 & 0.249 &    0.008 & $0.255\pm0.003$ &                 &                 &                 &                 \\
NGC\,1261  & 0.39 & 0.249 &    0.004 & $0.251\pm0.003$ &                 &                 &                 &                 \\
NGC\,5024  & 0.33 & 0.247 &    0.013 & $0.256\pm0.004$ & $0.100$ & $0.020$ & $0.113\pm0.018$ & $0.263\pm0.023$ \\
NGC\,5053  & 0.42 & 0.247 & $-0.002$ & $0.246\pm0.007$ & $0.116$ &                 & $0.116\pm0.014$ & $0.320\pm0.019$ \\
NGC\,5272  & 0.46 & 0.248 &    0.016 & $0.257\pm0.003$ & $0.188$ & $0.052$ & $0.216\pm0.020$ & $0.459\pm0.023$ \\
NGC\,5466  & 0.47 & 0.247 &    0.002 & $0.249\pm0.009$ & $0.103$ & $0.016$ & $0.111\pm0.020$ & $0.264\pm0.023$ \\
NGC\,5897  & 0.55 & 0.247 &          & $0.255\pm0.010$ &                 &                 &                 &                 \\
NGC\,5904  & 0.29 & 0.249 &    0.012 & $0.258\pm0.003$ & $0.176$ & $0.040$ & $0.204\pm0.022$ & $0.404\pm0.029$ \\
NGC\,6093  & 0.35 & 0.248 &    0.011 & $0.255\pm0.004$ & $0.156$ & $0.110$ & $0.228\pm0.025$ & $0.397\pm0.029$ \\
NGC\,6101  & 0.65 & 0.247 &    0.005 & $0.249\pm0.005$ & $0.110$ & $0.010$ & $0.114\pm0.016$ & $0.282\pm0.020$ \\
NGC\,6171  & 0.40 & 0.251 &    0.014 & $0.259\pm0.006$ & $0.230$ & $0.013$ & $0.238\pm0.027$ & $0.528\pm0.034$ \\
NGC\,6205  & 0.21 & 0.248 &    0.020 & $0.264\pm0.003$ & $0.210$ & $0.063$ & $0.260\pm0.021$ & $0.526\pm0.027$ \\
NGC\,6218  & 0.38 & 0.249 &    0.009 & $0.255\pm0.004$ & $0.223$ & $0.047$ & $0.252\pm0.027$ & $0.540\pm0.032$ \\
NGC\,6254  & 0.51 & 0.248 &    0.006 & $0.251\pm0.004$ & $0.206$ & $0.027$ & $0.219\pm0.022$ & $0.519\pm0.026$ \\
NGC\,6341  & 0.30 & 0.247 &    0.022 & $0.263\pm0.003$ & $0.053$ & $0.067$ & $0.100\pm0.021$ & $0.149\pm0.027$ \\
NGC\,6352  & 0.50 & 0.255 &    0.019 & $0.265\pm0.007$ & $0.256$ &                 & $0.256\pm0.039$ & $0.583\pm0.053$ \\
NGC\,6362  & 0.57 & 0.251 &    0.003 & $0.252\pm0.006$ & $0.213$ & $0.037$ & $0.229\pm0.026$ & $0.482\pm0.034$ \\
NGC\,6366  & 0.56 & 0.256 &    0.011 & $0.261\pm0.005$ & $0.273$ &                 & $0.273\pm0.041$ & $0.626\pm0.056$ \\
NGC\,6397  & 0.35 & 0.247 &    0.006 & $0.251\pm0.005$ & $0.136$ & $0.010$ & $0.143\pm0.018$ & $0.365\pm0.020$ \\
NGC\,6541  & 0.40 & 0.247 &    0.024 & $0.262\pm0.003$ & $0.170$ & $0.050$ & $0.200\pm0.022$ & $0.448\pm0.022$ \\
NGC\,6723  & 0.36 & 0.251 &    0.005 & $0.254\pm0.003$ & $0.180$ & $0.053$ & $0.214\pm0.024$ & $0.401\pm0.032$ \\
NGC\,6752  & 0.33 & 0.248 &    0.015 & $0.258\pm0.003$ & $0.216$ & $0.060$ & $0.256\pm0.023$ & $0.544\pm0.030$ \\
NGC\,6779  & 0.47 & 0.247 &    0.012 & $0.254\pm0.004$ & $0.140$ & $0.033$ & $0.157\pm0.018$ & $0.374\pm0.020$ \\
NGC\,6809  & 0.24 & 0.247 &    0.014 & $0.258\pm0.004$ & $0.140$ & $0.033$ & $0.165\pm0.023$ & $0.370\pm0.023$ \\
NGC\,6838  & 0.66 & 0.255 &    0.005 & $0.257\pm0.005$ & $0.210$ &                 & $0.210\pm0.018$ & $0.466\pm0.024$\\
NGC\,7099  & 0.38 & 0.247 &    0.015 & $0.257\pm0.005$ & $0.066$ & $0.017$ & $0.077\pm0.017$ & $0.193\pm0.019$ \\
\hline
\end{tabular}
\]
\end{table}

We select clusters with two main stellar generations, hereafter designated as 1G and 2G, respectively, with a primordial helium mass fraction $Y_\mathrm{1G}\approx0.25$ for 1G and a mild difference $\delta Y_\mathrm{2G,1G}$ between the generations \citep{milone2018}. The generations have a similar enrichment by $\alpha$ elements about $0.2<[\alpha/$Fe$]<0.4$ \citep{carretta2010,masseron2019,meszaros2020}, even in the most metal-rich clusters under consideration, e.g. $[\alpha/$Fe$]\approx0.2$ for NGC\,6352 \citep{feltzing2009}, $[\alpha/$Fe$]=0.28$ for NGC\,6366 \citep{puls2018}, and $[\alpha/$Fe$]=0.40$ \citep{carretta2010} or $0.32\pm0.08$ \citep{meszaros2020} for NGC\,6838.
We have tested fitting of NGC\,6352 and NGC\,6366 by the DSED isochrones with a lower $\alpha$-enhancement $[\alpha/$Fe$]=0.2$ (no such isochrones from BaSTI): this changes the parameter estimates insignificantly, within the declared uncertainties.

For our isochrone fitting of observed mix of the generations, we adopt the helium mass fraction $Y_\mathrm{mix}$ of the mix calculated as
\begin{equation}
\label{ymix}
Y_\mathrm{mix}=N_\mathrm{1G}/N_\mathrm{TOT}\cdot Y_\mathrm{1G}+(1-N_\mathrm{1G}/N_\mathrm{TOT})\cdot (Y_\mathrm{1G}+\delta Y_\mathrm{2G,1G}),
\end{equation}
where $N_\mathrm{1G}/N_\mathrm{TOT}$ is the fraction of 1G stars adopted from \citet{jang2025} or \citet{dondoglio2021}, $Y_\mathrm{1G}$ is stated by BaSTI or DSED for the cluster's $[$Fe$/$H$]$ and $[\alpha/$Fe$]$, and $\delta Y_\mathrm{2G,1G}$ is adopted from \citet{milone2018}.
The uncertainty of $Y_\mathrm{mix}$ is calculated from the uncertainties of the input arguments.
We adopt $Y_\mathrm{mix}=0.255\pm0.010$ for NGC\,5897 with no $\delta Y_\mathrm{2G,1G}$ estimates.
All these quantities are presented in Table~\ref{mix}
together with the mass loss $\mu_\mathrm{1G}$ for 1G (in solar mass), mass-loss difference $\delta \mu_\mathrm{2G,1G}$ between the generations, mass loss efficiency described by the free parameter $\eta$ in Reimers law \citep{reimers} -- all adopted from \citet[][hereafter TML20]{tailo2020}, as well as the mass loss $\mu_\mathrm{mix}$ for the mix of the generations calculated similarly to Equation~(\ref{ymix}):
\begin{equation}
\label{mlmix}
\mu_\mathrm{mix}=N_\mathrm{1G}/N_\mathrm{TOT}\cdot \mu_\mathrm{1G}+(1-N_\mathrm{1G}/N_\mathrm{TOT})\cdot (\mu_\mathrm{1G}+\delta \mu_\mathrm{2G,1G}).
\end{equation}

\begin{table*}
\def\baselinestretch{1}\normalsize\scriptsize
\caption[]{The parameters derived in our study: truncation radius $R_t$ (arcmin),
$[$Fe$/$H$]$ (dex), age (Gyr), distance $D$ (kpc), distance modulus DM (mag), $E(B-V)$ (mag), and apparent $V$-band distance modulus DMV (mag).
The total uncertainties are provided (see the text).
}
\label{estimates}
\[
\begin{tabular}{lccccccc}
\hline
\noalign{\smallskip}
Cluster    & $R_t$ & $[$Fe$/$H$]$ & Age   & $D$  & DM & $E(B-V)$ & DMV \\
\hline
\noalign{\smallskip}
NGC\,288   & 18 & $-1.25\pm0.05$ & $13.00\pm0.50$ &  $8.88\pm0.18$ & $14.74\pm0.04$ & $0.016\pm0.029$ & $14.80\pm0.07$ \\
NGC\,362   & 20 & $-1.25\pm0.05$ & $10.58\pm0.25$ &  $8.89\pm0.10$ & $14.75\pm0.02$ & $0.027\pm0.027$ & $14.84\pm0.07$ \\
NGC\,1261  & 16 & $-1.35\pm0.15$ & $10.50\pm0.50$ & $16.52\pm0.20$ & $16.09\pm0.03$ & $0.023\pm0.033$ & $16.17\pm0.12$ \\
NGC\,5024  & 15 & $-1.90\pm0.05$ & $12.67\pm0.25$ & $18.38\pm0.25$ & $16.32\pm0.03$ & $0.026\pm0.020$ & $16.41\pm0.06$ \\
NGC\,5053  & 10 & $-2.22\pm0.15$ & $12.42\pm0.50$ & $17.36\pm0.35$ & $16.20\pm0.04$ & $0.026\pm0.010$ & $16.29\pm0.03$ \\
NGC\,5272  & 23 & $-1.57\pm0.05$ & $11.50\pm0.50$ & $10.13\pm0.15$ & $15.03\pm0.03$ & $0.027\pm0.021$ & $15.12\pm0.08$ \\
NGC\,5466  & 17 & $-2.00\pm0.10$ & $12.00\pm0.25$ & $15.74\pm0.37$ & $15.98\pm0.05$ & $0.030\pm0.017$ & $16.09\pm0.05$ \\
NGC\,5897  & 10 & $-1.92\pm0.05$ & $12.75\pm0.25$ & $12.57\pm0.18$ & $15.50\pm0.03$ & $0.135\pm0.020$ & $15.97\pm0.05$ \\
NGC\,5904  & 35 & $-1.30\pm0.10$ & $11.58\pm0.25$ &  $7.26\pm0.21$ & $14.30\pm0.06$ & $0.040\pm0.028$ & $14.44\pm0.10$ \\
NGC\,6093  &  7 & $-1.70\pm0.05$ & $12.50\pm0.25$ & $10.46\pm0.15$ & $15.10\pm0.03$ & $0.212\pm0.026$ & $15.83\pm0.07$ \\
NGC\,6101  & 17 & $-1.98\pm0.10$ & $12.33\pm0.50$ & $14.21\pm0.35$ & $15.76\pm0.05$ & $0.132\pm0.021$ & $16.22\pm0.10$ \\
NGC\,6171  & 10 & $-1.02\pm0.10$ & $11.83\pm0.25$ &  $5.46\pm0.04$ & $13.68\pm0.01$ & $0.429\pm0.017$ & $15.18\pm0.07$ \\
NGC\,6205  & 27 & $-1.60\pm0.10$ & $12.75\pm0.75$ &  $7.40\pm0.03$ & $14.35\pm0.01$ & $0.026\pm0.023$ & $14.44\pm0.08$ \\
NGC\,6218  & 19 & $-1.25\pm0.05$ & $13.08\pm0.75$ &  $4.96\pm0.13$ & $13.48\pm0.06$ & $0.200\pm0.027$ & $14.17\pm0.10$ \\
NGC\,6254  & 19 & $-1.60\pm0.10$ & $13.00\pm0.50$ &  $5.11\pm0.09$ & $13.54\pm0.04$ & $0.270\pm0.020$ & $14.48\pm0.06$ \\
NGC\,6341  & 19 & $-2.25\pm0.05$ & $12.75\pm0.75$ &  $8.50\pm0.15$ & $14.65\pm0.04$ & $0.033\pm0.015$ & $14.76\pm0.06$ \\
NGC\,6352  & 10 & $-0.68\pm0.15$ & $12.50\pm0.50$ &  $5.30\pm0.09$ & $13.62\pm0.03$ & $0.290\pm0.050$ & $14.63\pm0.16$ \\
NGC\,6362  & 10 & $-1.03\pm0.15$ & $11.92\pm0.25$ &  $7.73\pm0.14$ & $14.44\pm0.04$ & $0.074\pm0.033$ & $14.70\pm0.12$ \\
NGC\,6366  & 15 & $-0.63\pm0.10$ & $11.33\pm0.25$ &  $3.30\pm0.06$ & $12.59\pm0.04$ & $0.749\pm0.031$ & $15.20\pm0.07$ \\
NGC\,6397  & 41 & $-1.92\pm0.10$ & $12.83\pm0.75$ &  $2.45\pm0.09$ & $11.95\pm0.08$ & $0.194\pm0.016$ & $12.62\pm0.07$ \\
NGC\,6541  & 11 & $-1.80\pm0.10$ & $13.17\pm0.25$ &  $7.56\pm0.07$ & $14.39\pm0.02$ & $0.132\pm0.020$ & $14.85\pm0.07$ \\
NGC\,6723  &  9 & $-1.07\pm0.15$ & $12.33\pm0.25$ &  $8.09\pm0.12$ & $14.54\pm0.03$ & $0.080\pm0.032$ & $14.82\pm0.10$ \\
NGC\,6752  & 38 & $-1.48\pm0.10$ & $13.25\pm0.75$ &  $4.03\pm0.09$ & $13.03\pm0.05$ & $0.061\pm0.026$ & $13.24\pm0.10$ \\
NGC\,6779  & 16 & $-2.00\pm0.10$ & $12.33\pm0.50$ & $10.70\pm0.25$ & $15.15\pm0.05$ & $0.254\pm0.012$ & $16.03\pm0.05$ \\
NGC\,6809  & 18 & $-1.83\pm0.05$ & $12.42\pm0.50$ &  $5.30\pm0.07$ & $13.62\pm0.03$ & $0.130\pm0.020$ & $14.07\pm0.06$ \\
NGC\,6838  & 10 & $-0.73\pm0.15$ & $12.17\pm0.50$ &  $3.99\pm0.04$ & $13.01\pm0.02$ & $0.240\pm0.034$ & $13.84\pm0.11$ \\
NGC\,7099  & 13 & $-2.10\pm0.15$ & $12.58\pm0.50$ &  $8.34\pm0.07$ & $14.60\pm0.02$ & $0.049\pm0.011$ & $14.78\pm0.03$ \\
\hline
\end{tabular}
\]
\end{table*}

\section{New isochrone fitting}
\label{newfitting}

For each cluster, we use the photometry in the $G_\mathrm{BP}$ (effective wavelength $\lambda_\mathrm{eff}=505$ nm) and $G_\mathrm{RP}$ ($\lambda_\mathrm{eff}=770$ nm) filters from {\it Gaia} DR3,
the $F606W$ ($\lambda_\mathrm{eff}=599$ nm) and $F814W$ ($\lambda_\mathrm{eff}=807$ nm) filters from the {\it HST} ACS survey,
and the $B$ ($\lambda_\mathrm{eff}=452$ nm) and $I$ ($\lambda_\mathrm{eff}=807$ nm) filters from \citetalias{stetson2019}.
Accordingly, we fit three CMDs for each cluster: $G_\mathrm{BP}-G_\mathrm{RP}$ vs $G_\mathrm{RP}$, $F606W-F814W$ vs $F814W$, and $B-I$ vs $I$.

The cleaning of the data sets is described in our previous papers and generally it follows the recommendations of the data set authors to select single star-like objects with photometric uncertainty better than several hundredths of a magnitude.
Anyway, the contribution of the photometric uncertainty to the total one must be much lower than the contribution of systematic uncertainties of models or observations, as discussed in our previous papers.

Our selection of the {\it Gaia} DR3 cluster members is described in our previous papers.
Briefly, we truncate the data sets at truncation radii $R_t$ where the cluster member count surface density decreases to its Galactic background.
Since our $R_t$ presented in Table~\ref{estimates} are lower than the tidal radii of the clusters presented, e.g. in the comprehensive cluster database of \citet[][hereafter BV21]{baumgardt2021}\footnote{url{https://people.smp.uq.edu.au/HolgerBaumgardt/globular/}}, we may lose few cluster members  beyond $R_t$, but this does not affect our results.
We retain only stars with appropriate {\it Gaia} DR3 parallaxes and proper motions (PMs).
Then we calculate the systemic PM components and coordinates of the cluster centres as medians for the members.
Since all the quantities used in the procedure of member selection are interrelated, we repeat it iteratively until negligible changes of the quantities.

\begin{figure*}
   \centering
   \includegraphics[width=14.5cm, angle=0]{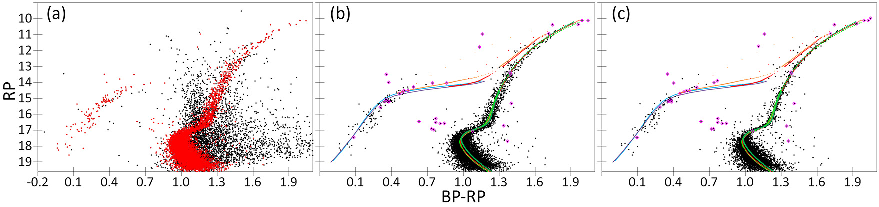}
   \caption{The {\it Gaia} $G_\mathrm{BP}-G_\mathrm{RP}$ vs $G_\mathrm{RP}$ CMD for NGC\,6254: (a) before the separation of cluster members (red symbols) from non-members (black symbols) but after the remaining cleaning of the data set, (b) after the selection of the cluster members but before the differential reddening correction, and (c) the final CMD.
Variable stars are shown in (b) and (c) by the magenta diamonds.
The isochrones for $Y\approx0.25$ from BaSTI (red), BaSTI ZAHB (purple), and DSED (green), isochrones for $Y=0.275$ from BaSTI (orange) and BaSTI ZAHB (blue), as well as isochrones for $Y=0.33$ from DSED (luminous green) are calculated with the best-fitting parameters from Table~\ref{cmds}.
}
\label{ngc6254gaia_}
\end{figure*}

\begin{figure*}
   \centering
   \includegraphics[width=10.0cm, angle=0]{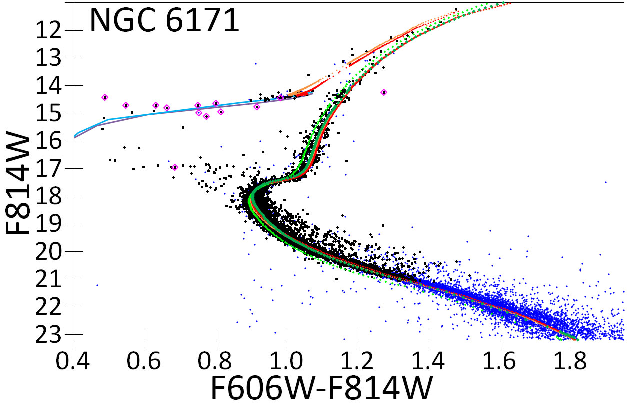}
   \caption{The \citetalias{nardiello2018} $F606W-F814W$ vs $F814W$ CMD for NGC\,6171 for cluster members with membership probability $>0.9$ (black symbols) and undefined membership probability $=-1$ (blue symbols).
Variable stars are shown by the magenta diamonds.
The isochrones for $Y\approx0.25$ from BaSTI (red), BaSTI ZAHB (purple), and DSED (green), isochrones for $Y=0.275$ from BaSTI (orange) and BaSTI ZAHB (blue), as well as isochrones for $Y=0.33$ from DSED (luminous green) are calculated with the best-fitting parameters from Table~\ref{cmds}.
It is worth noting that the best-fit BaSTI ZAHB exactly fits the faintest red HB stars and ignores RR~Lyrae variables.
Interestingly, the combination of the F606W and F814W filters with the RR~Lyrae SEDs make some of these variable stars redder than the RGB as seen in this figure for the star NGC\,6171 V8 at $F606W-F814W=1.27$ and $F814W=14.24$ mag.}
\label{ngc6171hst_}
\end{figure*}

\begin{figure*}
   \centering
   \includegraphics[width=10.0cm, angle=0]{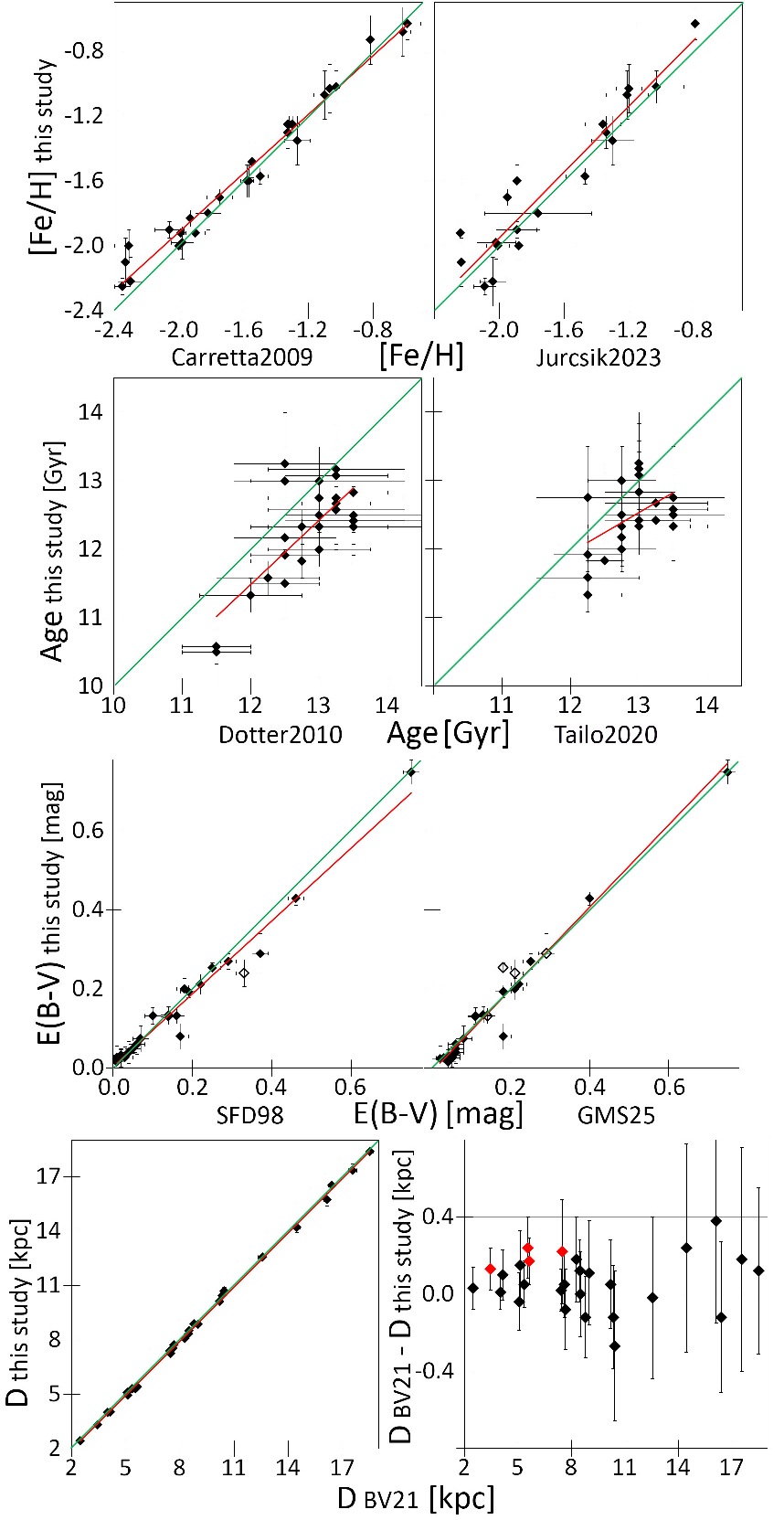}
   \caption{Comparison of our estimates of $[$Fe$/$H$]$ with those from \citet{carretta2009} and \citet{jurcsik2023}, of age with those from \citet{dotter2010} and \citetalias{tailo2020}, of $E(B-V)$ with those from \citetalias{sfd98} and \citetalias{gms2025}, and of $D$ with those from \citetalias{baumgardt2021}.
   The green line represent the one-to-one correspondence, while the red line does the best root mean square relation.
   The empty diamonds indicate clusters, for which $E(B-V)$ may be over- or underestimated by \citetalias{sfd98} and \citetalias{gms2025}, respectively (see the text).
   The red diamonds indicate four clusters with a large difference between our and \citetalias{baumgardt2021} $D$ estimates exceeding the sum of their uncertainties.}
\label{comparison}
\end{figure*}

\begin{figure*}
   \centering
   \includegraphics[width=12.0cm, angle=0]{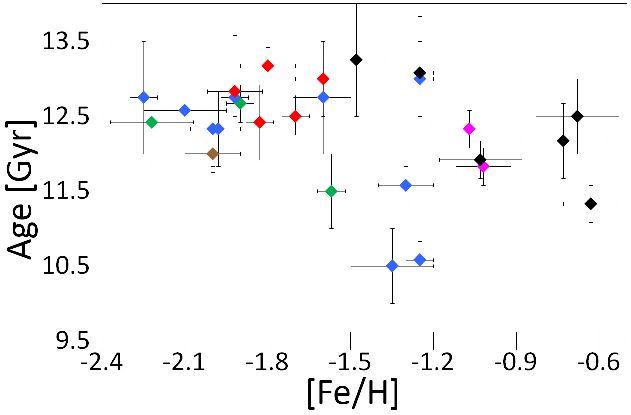}
   \caption{$[$Fe$/$H$]$ vs age for our sample of the clusters.
   Clusters of different association are marked by different colours (see the text).
}
\label{feh_age}
\end{figure*}

Since many clusters are strongly contaminated by field stars, the selection of cluster members is very important as illustrated in Fig.~\ref{ngc6254gaia_} for the {\it Gaia} CMD of NGC\,6254. It is seen that before the member selection, the CMD is so heavily contaminated, mainly by the Sagittarius dwarf galaxy members behind NGC\,6254, that becomes difficult to fit an isochrone to the data and, hence, to derive accurate estimates of the cluster parameters.
Accordingly, we cross-identify the {\it Gaia} DR3 and \citetalias{stetson2019} data sets and consider only the {\it Gaia} cluster members from the latter.\footnote{Since the {\it Gaia} photometry is not important when we fit the \citetalias{stetson2019} CMDs, we select the {\it Gaia} cluster members among the \citetalias{stetson2019} data sets using only their {\it Gaia} astrometry irrespective of their {\it Gaia} photometry. Therefore, for some clusters we consider more {\it Gaia} cluster members in the \citetalias{stetson2019} CMD than in the corresponding {\it Gaia} CMD.}

Table~\ref{systemic} presents our estimates of the systemic PM components for 13 newly fitted clusters\footnote{Our identification of variable stars in nine clusters from \citetalias{ngc6362}--\citetalias{ngc5024} and redo of their isochrone fitting in this study does not affect their systemic PMs and {\it Gaia} DR3 parallaxes.
Their systemic PMs can be found in the corresponding papers, as those for five clusters from \citetalias{paper7}, while their parallaxes are shown in Table~\ref{parallax} for further comparisons.}
in comparison with those from \citet[][hereafter VB21]{vasiliev2021}, which are also derived from the {\it Gaia} DR3 PMs but using a different approach.
The total (statistic plus systematic) uncertainty of all these estimates is dominated by the same systematics, which cannot be better than 0.02 mas\,yr$^{-1}$, as shown by \citetalias{vasiliev2021}, and, hence, is the same regardless of approach.
Table~\ref{systemic} shows that all the estimates agree within the uncertainties.

\begin{table*}
\def\baselinestretch{1}\normalsize\normalsize
\caption[]{The cluster systemic PM components (mas\,yr$^{-1}$) with their total (statistic and systematic) uncertainties for 13 newly fitted clusters.
}
\label{systemic}
\[
\begin{tabular}{llcc}
\hline
\noalign{\smallskip}
Cluster & Source & $\mu_{\alpha}\cos(\delta)$ & $\mu_{\delta}$ \\
\hline
\noalign{\smallskip}
NGC\,1261 & This study                & $1.602\pm0.024$ & $-2.064\pm0.025$ \\
NGC\,1261 & \citetalias{vasiliev2021} & $1.596\pm0.024$ & $-2.064\pm0.025$ \\
\noalign{\smallskip}
NGC\,5897 & This study                & $-5.428\pm0.025$ & $-3.391\pm0.025$ \\
NGC\,5897 & \citetalias{vasiliev2021} & $-5.422\pm0.025$ & $-3.392\pm0.025$ \\
\noalign{\smallskip}
NGC\,6093 & This study                & $-2.939\pm0.027$ & $-5.566\pm0.026$ \\
NGC\,6093 & \citetalias{vasiliev2021} & $-2.935\pm0.027$ & $-5.577\pm0.026$ \\
\noalign{\smallskip}
NGC\,6101 & This study                & $1.751\pm0.025$ & $-0.264\pm0.024$ \\
NGC\,6101 & \citetalias{vasiliev2021} & $1.755\pm0.025$ & $-0.258\pm0.024$ \\
\noalign{\smallskip}
NGC\,6171 & This study                & $-1.934\pm0.025$ & $-5.976\pm0.025$ \\
NGC\,6171 & \citetalias{vasiliev2021} & $-1.939\pm0.025$ & $-5.979\pm0.025$ \\
\noalign{\smallskip}
NGC\,6254 & This study                & $-4.755\pm0.024$ & $-6.609\pm0.024$ \\
NGC\,6254 & \citetalias{vasiliev2021} & $-4.757\pm0.024$ & $-6.596\pm0.024$ \\
\noalign{\smallskip}
NGC\,6341 & This study                & $-4.931\pm0.025$ & $-0.632\pm0.023$ \\
NGC\,6341 & \citetalias{vasiliev2021} & $-4.936\pm0.025$ & $-0.625\pm0.023$ \\
\noalign{\smallskip}
NGC\,6352 & This study                & $-2.162\pm0.024$ & $-4.445\pm0.025$ \\
NGC\,6352 & \citetalias{vasiliev2021} & $-2.157\pm0.024$ & $-4.446\pm0.025$ \\
\noalign{\smallskip}
NGC\,6366 & This study                & $-0.337\pm0.025$ & $-5.168\pm0.024$ \\
NGC\,6366 & \citetalias{vasiliev2021} & $-0.332\pm0.025$ & $-5.159\pm0.024$ \\
\noalign{\smallskip}
NGC\,6541 & This study                & $0.271\pm0.025$ & $-8.838\pm0.025$ \\
NGC\,6541 & \citetalias{vasiliev2021} & $0.287\pm0.025$ & $-8.847\pm0.025$ \\
\noalign{\smallskip}
NGC\,6752 & This study                & $-3.160\pm0.023$ & $-4.037\pm0.022$ \\
NGC\,6752 & \citetalias{vasiliev2021} & $-3.162\pm0.023$ & $-4.028\pm0.022$ \\
\noalign{\smallskip}
NGC\,6779 & This study                & $-2.011\pm0.025$ & $1.606\pm0.025$ \\
NGC\,6779 & \citetalias{vasiliev2021} & $-2.018\pm0.025$ & $1.617\pm0.025$ \\
\noalign{\smallskip}
NGC\,6838 & This study                & $-3.402\pm0.025$ & $-2.669\pm0.025$ \\
NGC\,6838 & \citetalias{vasiliev2021} & $-3.416\pm0.025$ & $-2.654\pm0.025$ \\
\hline
\end{tabular}
\]
\end{table*}
\begin{table*}
\def\baselinestretch{1}\normalsize\normalsize
\caption[]{Parallax estimates (mas), obtained by various methods, with their total (statistic and systematic) uncertainties for 22 clusters fitted in this study.
}
\label{parallax}
\[
\begin{tabular}{lcccc}
\hline
\noalign{\smallskip}
   & \citetalias{vasiliev2021} & \citetalias{baumgardt2021} & \multicolumn{2}{c}{This study} \\
Cluster   & {\it Gaia} DR3 astrometry & Various methods & {\it Gaia} DR3 astrometry & Isochrone fitting \\
\hline
\noalign{\smallskip}
NGC\,1261 & $0.064\pm0.011$ & $0.061\pm0.001$ & $0.065\pm0.011$ & $0.061\pm0.001$ \\ 
NGC\,5024 & $0.064\pm0.011$ & $0.054\pm0.001$ & $0.066\pm0.011$ & $0.054\pm0.001$ \\ 
NGC\,5053 & $0.047\pm0.011$ & $0.057\pm0.001$ & $0.041\pm0.015$ & $0.058\pm0.001$ \\ 
NGC\,5272 & $0.106\pm0.010$ & $0.098\pm0.001$ & $0.110\pm0.010$ & $0.099\pm0.001$ \\ 
NGC\,5466 & $0.053\pm0.011$ & $0.062\pm0.001$ & $0.062\pm0.011$ & $0.064\pm0.001$ \\ 
NGC\,5897 & $0.101\pm0.011$ & $0.080\pm0.002$ & $0.096\pm0.011$ & $0.080\pm0.001$ \\ 
NGC\,6093 & $0.098\pm0.011$ & $0.097\pm0.001$ & $0.095\pm0.011$ & $0.096\pm0.001$ \\ 
NGC\,6101 & $0.080\pm0.011$ & $0.069\pm0.001$ & $0.082\pm0.011$ & $0.070\pm0.001$ \\ 
NGC\,6171 & $0.190\pm0.011$ & $0.178\pm0.002$ & $0.178\pm0.011$ & $0.185\pm0.003$ \\ 
NGC\,6254 & $0.193\pm0.010$ & $0.197\pm0.002$ & $0.183\pm0.011$ & $0.196\pm0.003$ \\ 
NGC\,6341 & $0.108\pm0.010$ & $0.118\pm0.001$ & $0.114\pm0.011$ & $0.118\pm0.002$ \\ 
NGC\,6352 & $0.186\pm0.011$ & $0.180\pm0.002$ & $0.186\pm0.011$ & $0.189\pm0.003$ \\ 
NGC\,6362 & $0.132\pm0.011$ & $0.131\pm0.001$ & $0.123\pm0.011$ & $0.129\pm0.002$ \\ 
NGC\,6366 & $0.281\pm0.011$ & $0.290\pm0.004$ & $0.275\pm0.011$ & $0.302\pm0.007$ \\ 
NGC\,6397 & $0.414\pm0.010$ & $0.403\pm0.003$ & $0.416\pm0.010$ & $0.408\pm0.012$ \\ 
NGC\,6541 & $0.142\pm0.011$ & $0.131\pm0.002$ & $0.133\pm0.011$ & $0.132\pm0.001$ \\ 
NGC\,6723 & $0.129\pm0.011$ & $0.121\pm0.001$ & $0.123\pm0.011$ & $0.124\pm0.001$ \\ 
NGC\,6752 & $0.251\pm0.010$ & $0.242\pm0.002$ & $0.250\pm0.011$ & $0.248\pm0.004$ \\ 
NGC\,6779 & $0.087\pm0.011$ & $0.096\pm0.001$ & $0.094\pm0.011$ & $0.093\pm0.002$ \\ 
NGC\,6809 & $0.206\pm0.010$ & $0.187\pm0.002$ & $0.203\pm0.010$ & $0.189\pm0.003$ \\ 
NGC\,6838 & $0.247\pm0.011$ & $0.250\pm0.003$ & $0.251\pm0.011$ & $0.251\pm0.003$ \\ 
NGC\,7099 & $0.132\pm0.011$ & $0.118\pm0.001$ & $0.119\pm0.011$ & $0.120\pm0.001$ \\ 
\hline
\end{tabular}
\]
\end{table*}

The parallaxes of cluster members are corrected for the parallax zero-point \citep{lindegren2021}.
The median corrected parallaxes for 13 newly fitted clusters together with those for 9 clusters from \citetalias{ngc6362}--\citetalias{ngc5024} are presented in Table~\ref{parallax} for further comparison with other estimates.
We adopt the total uncertainty of the derived parallaxes, as determined by \citetalias{vasiliev2021}, not less than 0.01 mas.

The authors of the \citetalias{nardiello2018} data sets use dedicated {\it HST} PMs for almost all rather bright stars (nearly $F606W<22$~mag, i.e. nearly Johnson $V<22$~mag) in the cluster fields in order to assign membership probability to them.
We eliminate stars with the membership probability between 0\% and 90\%.
Fig.~\ref{ngc6171hst_} presents an example of the \citetalias{nardiello2018} CMD for NGC\,6171, where the stars with the membership probability $>90\%$ are shown by the black symbols.
However, fainter stars and some bright stars observed by {\it HST} have no PMs and, hence, no membership probability estimation. The authors of the {\it HST} data sets assign them with the membership probability $=-1$.
Such stars are shown in Fig.~\ref{ngc6171hst_} by the blue symbols.
Members dominate in the \citetalias{nardiello2018} data sets, since these data sets cover only few central arcminutes of the cluster fields.
Fig.~\ref{ngc6171hst_} shows that even most stars with indeterminate membership seem to be members judging by their position in the CMD.
This is the same for all the clusters, except NGC\,6397.\footnote{NGC\,6397 is so strongly contaminated by the Sagittarius dwarf galaxy (see \citetalias{ngc6397}) that it demonstrates too many non-members among its stars with indeterminate membership as seen in its \citetalias{nardiello2018} CMD presented in Appendix~\ref{addcmds}. Hence, for NGC\,6397 we use only stars with the membership probability $>90\%$ in our isochrone fitting of the \citetalias{nardiello2018} CMD.}
Therefore, we retain stars with indeterminate membership for all the clusters except NGC\,6397, since faint MS stars among them are important for our determination of $[$Fe$/$H$]$.
It is worth noting that $[$Fe$/$H$]$ can be best determined from fitting of the bright RGB or faint MS stars. We use the former for the determination of the best-fit $[$Fe$/$H$]$ from the {\it Gaia} and \citetalias{stetson2019} CMDs due to their incomplete faint MS, while the latter are used for the \citetalias{nardiello2018} CMDs due to their sparsely populated bright RGB.
It is remarkable that our $[$Fe$/$H$]$ estimates derived for the different data sets agree, as seen in Table~\ref{cmds} in Appendix~\ref{best}.

We correct differential reddening (DR) in all CMDs with enough (nearly $>3000$) stars using the method of \citet{bonatto2013}.
Briefly, we divide the cluster area into a grid of cells, with higher angular resolution in regions of the area containing more stars.
Then, the stellar density Hess diagram (including photometric errors) of each cell is fitted to such diagram, averaged over the entire cluster area, by its shift along the reddening vector. This shift is then converted into DR in the cell. The same DR correction is applied to all stars in one cell.

Usually, DR in the wide areas covered by the {\it Gaia} or \citetalias{stetson2019} data sets is much higher than DR in the few central arcminutes covered by the \citetalias{nardiello2018} data sets (e.g. see the NGC\,6723 DR in \citetalias{ngc6362}).

Among all 27 clusters, only NGC\,6093, NGC\,6171, NGC\,6254, NGC\,6352, NGC\,6366, NGC\,6397 (see \citetalias{ngc6397}), NGC\,6723 (see \citetalias{ngc6362}), and NGC\,6838 demonstrate a significant DR, that is, with the difference $\Delta E(B-V)>0.05$~mag between the 98$^\mathrm{th}$ and the second percentile of the DR distributions.
This is in line with the DR estimates for the same clusters obtained by \citet{alonso2012,jang2022,pancino2024}.

Fig.~\ref{ngc6254gaia_} presents an example of changes in a CMD after the DR correction: the scatter of the NGC\,6254 members around the best-fitting isochrones in its {\it Gaia} CMD reduces making the parameter estimates more accurate.

To fit CMDs, we use the $\alpha$--enhanced $[\alpha/$Fe$]=+0.4$ isochrones from BaSTI with initial solar $Z=0.0172$ and $Y=0.2695$, overshooting, diffusion, mass loss efficiency described by $\eta=0.3$ in Reimers law,
as well as the $\alpha$--enhanced $[\alpha/$Fe$]=+0.4$ DSED isochrones with solar $Z=0.0189$ and no mass loss.
Also we use the BaSTI extended set of zero-age horizontal branch (ZAHB) predictions with a stochastic mass loss before the HB.
To interpolate the isochrones for the adopted $Y_\mathrm{mix}$ (see Table~\ref{mix}), we use the BaSTI isochrones for $Y\approx0.25$ and $0.275$, while the DSED ones for $Y\approx0.25$ and $0.33$.

The accurate selection of cluster members and DR correction make the distribution of stars in our CMDs well defined.
Hence, as in \citetalias{ngc6397}, \citetalias{ngc5024}, and \citetalias{paper7}, we fit isochrones directly to the bulk of cluster members, without calculating a fiducial sequence.
To balance the contribution of different CMD domains, we assign a weight to each data point, which is inversely proportional to the number of stars of a given magnitude.
We exclude some stars from the fitting:
the extremely blue HB (with effective temperature $T_\mathrm{eff}>9000$ K),
RR~Lyrae stars, some other variable stars, and blue stragglers.

We evaluate numerous sets of the parameters ($[$Fe$/$H$]$, distance from the Sun $D$, reddening, and age) in their 4-dimensional space with their steps of 0.1 dex, 0.01 kpc, 0.001 mag, and 0.5 Gyr, respectively, for each CMD-model pair.
For each parameter set, we calculate the sum of the squares of the residuals between the isochrones and the data points.
The set with the minimal sum is the best solution, which is given for each CMD in Table~\ref{cmds} of Appendix~\ref{best}.
Two examples of CMDs with the best solutions are presented in Figs~\ref{ngc6254gaia_} and \ref{ngc6171hst_}, whereas all the best-fitted CMDs are provided in Appendix~\ref{addcmds}.

Our final estimates of $[$Fe$/$H$]$, age, distance $D$, distance modulus DM$\equiv(m-M)_0$, $E(B-V)$,
\footnote{To calculate an $E(B-V)$ estimate from any other extinction or reddening estimate, we use extinction coefficients from \citet{casagrande2014,casagrande2018a,casagrande2018b} or \citet{ccm89} with the observed extinction-to-reddening ratio $R_\mathrm{V}$ depending on the intrinsic spectral energy distribution (SED) of the cluster members under consideration \citep{casagrande2014}.
For the metal-poor and cool members of the clusters with typical effective temperature $T_\mathrm{eff}\approx5400$~K, $R_\mathrm{V}\approx3.4$. This value differs from $R_\mathrm{V}\equiv A_\mathrm{V}/E(B-V)=3.1$ defined for early type MS stars due to their different SED.}
and apparent $V$-band distance modulus DMV$\equiv(m-M)_\mathrm{V}$ are presented in Table~\ref{estimates}. They are calculated as the averages of six best-fitting estimates (three data sets fitted by two models).

In the case of few data sets and models, which may be systematically different, half the range seems to be a reasonable conservative upper limit of their total uncertainty.
Accordingly, we adopt the total uncertainty of a parameter for a cluster as half the range of all six estimates, but not lower than 0.05 dex, 0.25 Gyr, 0.01 mag, and 0.01 mag for $[$Fe$/$H$]$, age, DM, and $E(B-V)$, respectively, and corresponding lower limits for the uncertainties of $D$ and DMV.

Our $[$Fe$/$H$]$, age, $D$, and $E(B-V)$ estimates are compared in Figs~\ref{comparison} with those from Table~\ref{properties} and from the literature.

Our $[$Fe$/$H$]$ estimates agree with the photometric estimates from \citet{jurcsik2023},\footnote{If \citet{jurcsik2023} made only one observation for a cluster, we adopt its uncertainty as $\sigma([$Fe$/$H$])=0.35$.} based on their RR~Lyrae calibration, and spectroscopic estimates from \citet{carretta2009}, except a slight (by $\Delta[$Fe$/$H$]=0.08$) systematic overestimation of all our $[$Fe$/$H$]$ w.r.t. the former and a considerable overestimation of low $[$Fe$/$H$]<-2$ for four clusters w.r.t. the latter.
We have discussed in \citetalias{ngc5024} the discrepancy between spectroscopic and photometric estimates of $[$Fe$/$H$]$, especially for metal-poor clusters, as well as between various spectroscopic estimates themselves, such as between systematically lower and higher estimates from \citet{carretta2009} and \citet{meszaros2020}, respectively.
In line with that discussion, it seems that all the estimates under consideration, both ours and from the literature, support the arguments of \citet{mucciarelli2020} in favor of higher photometric and against lower spectroscopic metallicity estimates for metal-poor ($[$Fe$/$H$]<-2$) globular clusters.

Our age estimates are systematically lower than those from both \citet{dotter2010} (by about 0.6 Gyr) and \citetalias{tailo2020} (by about 0.4 Gyr). However, a relative age scale is more important \citep{catelan2018}. In this context, our age estimates can contribute to the formation of such a scale.

Our $E(B-V)$ estimates demonstrate a good agreement with the extinction/reddening maps of \citet[][hereafter SFD98]{sfd98} and \citet[][hereafter GMS25]{gms2025}.
The outliers marked in Fig.~\ref{comparison} by empty diamonds are
NGC\,6838 with a significant amount of background dust taken into account by \citetalias{sfd98}, which, hence, may overestimate its real $E(B-V)$ and
four clusters beyond the applicability space of \citetalias{gms2025}, which, hence, may underestimate their real $E(B-V)$.
Other outliers are NGC\,6171, NGC\,6352, and NGC\,6723 with their strong and hardly predictable DR (NGC\,6723 is affected by the adjacent foreground Corona Australis cloud complex as discussed in \citetalias{ngc6362}).

Our distance estimates agree with those from \citetalias{baumgardt2021} within their combined uncertainties, except four outliers (NGC\,5904, NGC\,6171, NGC\,6352, NGC\,6366) marked in Fig.~\ref{comparison} by the red diamonds. Since usually a theoretical ZAHB is fitted to a fainter bound of observed HB stars, it is more probable to bias this ZAHB to a fainter magnitude (resulting in an overestimation of distance) due to a strong DR, contamination by non-members, involvement of RR~Lyrae observed at random phase, or overestimation of the adopted $Y_\mathrm{mix}$ than to bias it to a brighter magnitude (resulting in an underestimation of distance) due to ignoring of the evolution of the HB stars from the ZAHB or underestimation of the adopted $Y_\mathrm{mix}$.
Indeed, the observed HBs of all outliers are highly contaminated by non-members before our cleaning and, moreover, among these outliers, NGC\,6352 and NGC\,6366 are affected by their high DR, while NGC\,5904 and NGC\,6171 contain many RR~Lyrae.
Since the data sets used in this study are corrected for the DR, contamination, and impact of variable stars (e.g. see Figs~\ref{ngc6254gaia_} and \ref{ngc6171hst_}) better than some data sets compiled by \citetalias{baumgardt2021}, we believe that our $D$ estimates are rather accurate w.r.t. those from \citetalias{baumgardt2021}.

We convert our new best-fitting $D$ and those of \citetalias{baumgardt2021}, along with their uncertainties, into parallaxes and their corresponding uncertainties for their comparison in Table~\ref{parallax} with the {\it Gaia} DR3 parallaxes obtained by us and \citetalias{vasiliev2021} (see such a comparison for the remaining five clusters in \citetalias{paper7}).
Table~\ref{parallax} shows that all the parallax estimates are consistent, with only four exceptions:
both trigonometric parallaxes for NGC\,5897, NGC\,6366, and NGC\,6809 significantly differ from both non-trigonometric estimates, while the \citetalias{vasiliev2021} parallax for NGC\,7099 significantly differs from the remaining estimates.
Since our and \citetalias{baumgardt2021} $D$ estimates, obtained by different methods, agree, this discrepancy may indicate an unaccounted systematics in the {\it Gaia} DR3 parallaxes.
It is worth noting, that the total uncertainty of any {\it Gaia} DR3 trigonometric parallax cannot be better than 0.01~mas \citepalias{vasiliev2021}, while the total uncertainty of parallax, derived in isochrone fitting, is nearly proportional to the parallax itself (see Table~\ref{parallax}).
Hence, for all the clusters, except NGC\,6397 (and any cluster with $D>3$ kpc), the {\it Gaia} DR3 parallaxes are less precise than those from our isochrone fitting.

\section{Relations between cluster parameters}
\label{relations}

Among the four derived parameters, $[$Fe$/$H$]$ and age are interesting in their relations to each other and to orbital, structural, and other parameters of the clusters.
We consider the adopted $Y_\mathrm{1G}$, $\delta Y_\mathrm{2G,1G}$, $Y_\mathrm{2G}$, and $N_\mathrm{1G}/N_\mathrm{TOT}$, mass loss estimates $\mu_\mathrm{1G}$ and $\mu_\mathrm{2G}$ from \citetalias{tailo2020}, cluster mass $M$ and core density from the database of \citet{baumgardt2021}, and estimates of the orbital parameters [apogalactic and perigalactic distance, eccentricity, inclination (or modified inclination defined between $0^{\circ}$ and $90^{\circ}$ as the angle of the orbit to the Galactic plane irrespective to the direction of the rotation), period, and time after the last crossing of the Galactic disk] from \citet{bajkova2022}.
We do not discuss the relations between the orbital parameters themselves, since this is done elsewhere in the literature.

Our consideration of core density as an important parameter requires the following comments.
In \citetalias{ngc5024} and \citetalias{paper7} we suggest that the loss of low-mass HB stars is an important factor of the HB morphology.
The lowest-mass (i.e. the bluest and the hottest) HB stars can be lost, among other low-mass cluster members, due to mass segregation, tidal shock in a rapidly changing Galactic potential during Galactic disk crossing, and other processes during cluster evolution \citep{odenkirchen2004,sollima2017}.
One should expect a strong loss of low-mass members in core-collapse and loose clusters, as well as in clusters with short Galactic period, low orbital inclination, or recent disk crossing.
Core-collapse and loose clusters may demonstrate the highest and lowest core density, respectively.
Indeed, among the clusters under consideration, NGC\,5053, NGC\,5466 (both analyzed in \citetalias{ngc5024}), and NGC\,6101 are loose clusters with their core density lower than 35 solar masses per cubic pc (see the \citetalias{baumgardt2021} data base).
On the contrary, NGC\,6093, NGC\,6341, NGC\,6397, NGC\,6541, NGC\,6752, and NGC\,7099 have core density higher than 50,000 solar masses per cubic pc.
Among them, NGC\,6397, NGC\,6752, and NGC\,7099 are known core-collapse clusters, NGC\,6541 is a core-collapse candidate.
We suggest that very high core density of NGC\,6093 and NGC\,6341 also allows us to consider them as core-collapse candidates.
Note that NGC\,362 and NGC\,6723 may be core-collapse clusters (see the \citetalias{baumgardt2021} data base), but they have rather low core density.
This may be due to very young age of NGC\,362, while the reason for NGC\,6723 is unknown.

Different origin, association, and evolution of the clusters may influence the relations between their parameters \citep{callingham2022,massari2025}.
Therefore, clusters of different origin and association are marked in the following figures with different colours.
Clusters associated with the disk, bulge, and low energy are marked by the black, magenta, and red symbols, respectively.
The status of the low energy subsample is uncertain: it is considered as accreted from the Kraken galaxy by \citet{kruijssen2019,kruijssen2020} or as in-situ by \citet{boldrini2025}, see also \citet{callingham2022}.
Moreover, NGC\,6093 belongs to either bulge \citep{callingham2022} or low energy \citep{massari2025} subsample,
NGC\,6397 and NGC\,6752 belong to either Kraken (low energy) \citep{callingham2022} or disk \citep{massari2025}.
Anyway, at least two subsamples should be considered as eight in-situ clusters: those associated with the disk
(NGC\,6218, NGC\,6352, NGC\,6362, NGC\,6366, NGC\,6752, NGC\,6838) and bulge (NGC\,6171 and NGC\,6723).
We associate NGC\,6397 with the low energy subsample following its similarities considered below.
Ex-situ (accreted) clusters associated with the Helmi stream, Gaia-Sausage-Enceladus progenitor, and Sequoia progenitor are marked by green, blue, and brown symbols, respectively.
The origin and association are taken from \citet[][hereafter CARMA]{aguado2025},\footnote{\url{https://www.oas.inaf.it/en/research/m2-en/carma-en/}}
while from \citet{boldrini2025,massari2025} for controversial cases.

Fig.~\ref{feh_age} presents the relation between the derived age and metallicity of the clusters (AMR relation). This figure reproduces a well-known pattern seen, for example, in figure 3 of \citet{kruijssen2020}, figure 9 of \citet{callingham2022}, and figure 3 of \citet{ceccarelli2025}.
In particular, history of Galactic globular clusters explains why the lowest metallicity clusters and the youngest clusters of moderate metallicity are ex-situ, while the most metal-rich clusters are in-situ and, furthermore, why the in-situ clusters demonstrate lower orbital inclinations, smaller apogalactic distances, intermediate perigalactic distances, shorter periods and, hence, smaller eccentricities as shown in Fig.~\ref{amr_orbit} in Appendix~\ref{more}.

\begin{figure}
\centering
\includegraphics[width=7.0cm, angle=0]{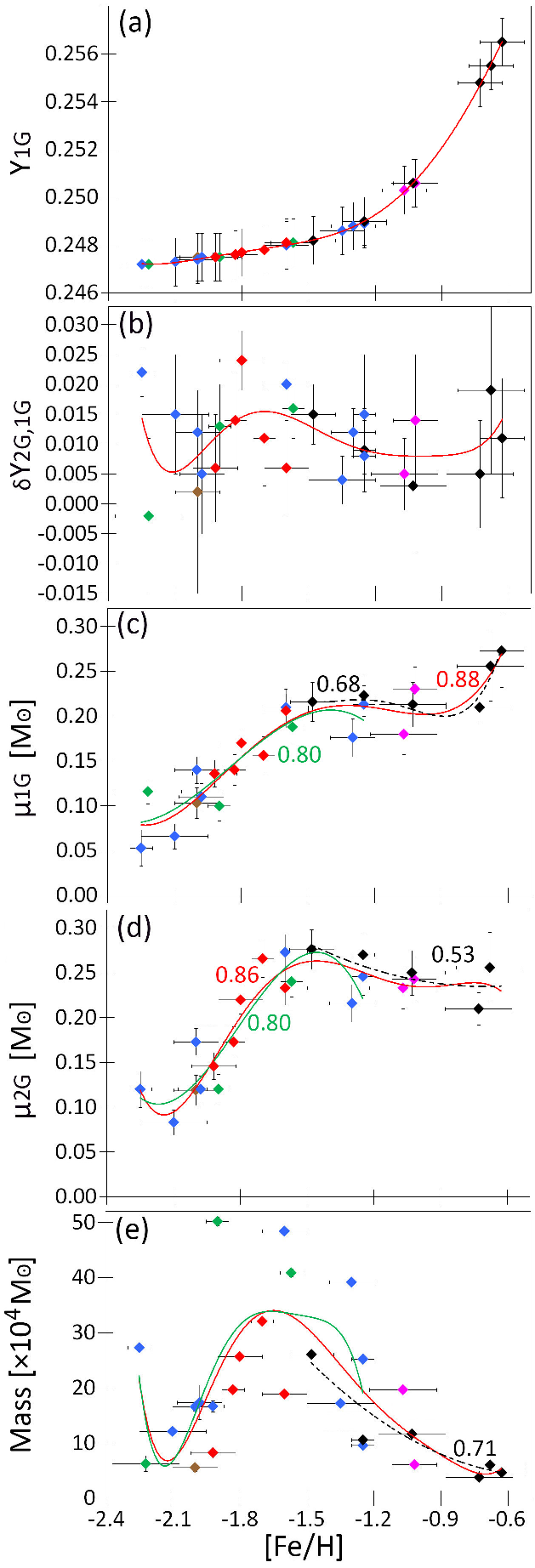}
\caption{$[$Fe$/$H$]$ vs $Y_\mathrm{1G}$, $\delta Y_\mathrm{2G,1G}$, $\mu_\mathrm{1G}$, $\mu_\mathrm{2G}$, and mass of the clusters.
The black dotted, green solid, and red solid curves show polynomial approximations for the in-situ, accreted, and all clusters, respectively.
The coefficients of determination $>0.5$ are marked near the polynomial curves by the same colour.
Clusters of different association are marked by different colours (see the text).
}
\label{feh}
\end{figure}

\begin{figure*}
\centering
\includegraphics[width=14.5cm, angle=0]{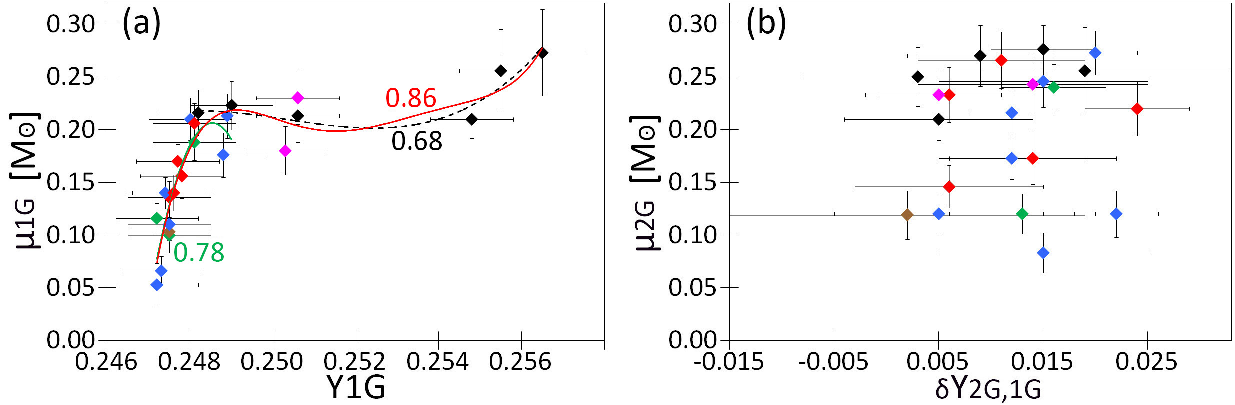}
\caption{The same as Fig.~\ref{feh} but for $Y_\mathrm{1G}$ vs $\mu_\mathrm{1G}$ and $\delta Y_\mathrm{2G,1G}$ vs $\mu_\mathrm{2G}$.}
\label{y_ml}
\end{figure*}

\begin{figure*}
\centering
\includegraphics[width=14.5cm, angle=0]{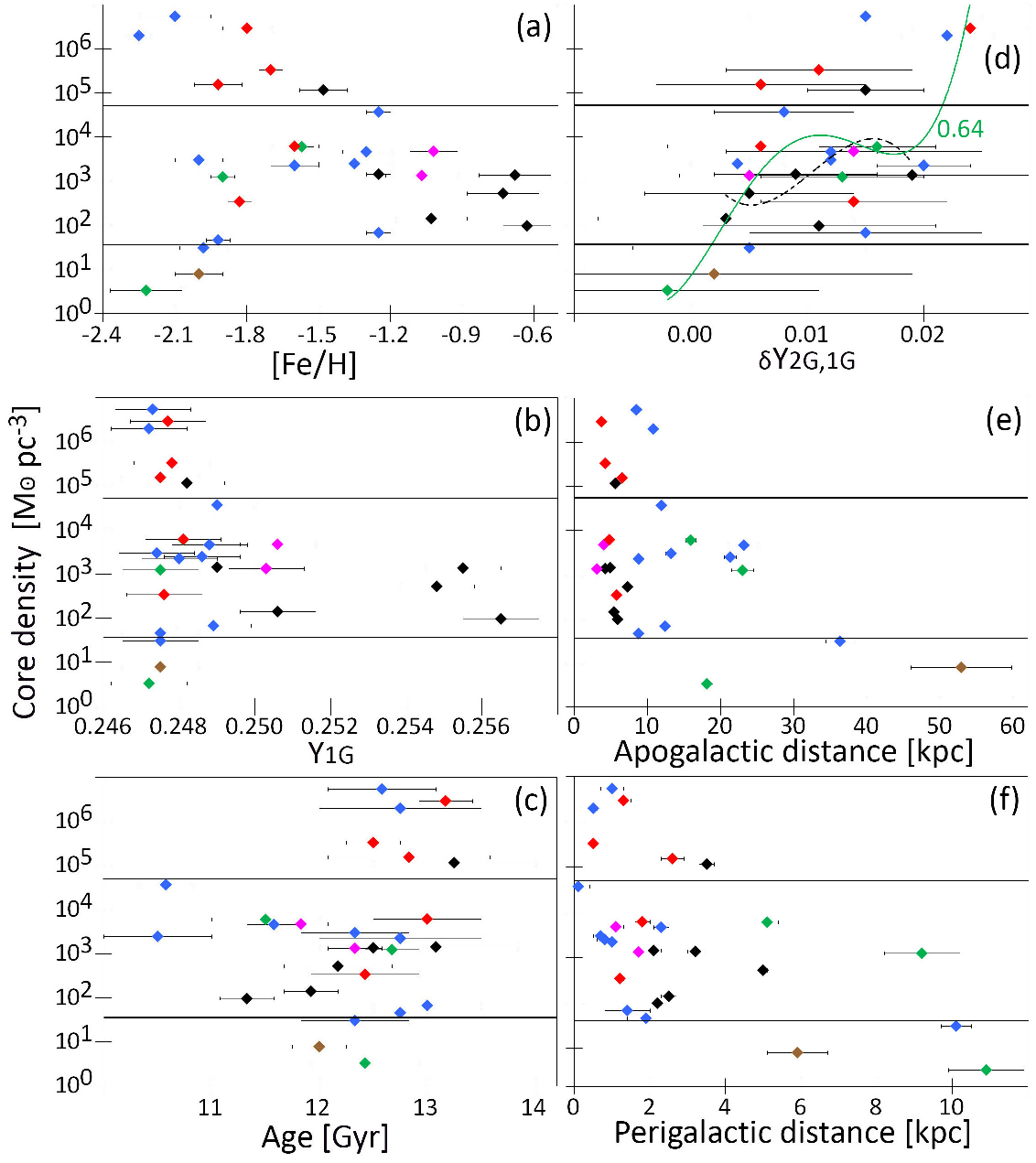}
\caption{The same as Fig.~\ref{feh} but for core density vs some parameters.
The black solid lines separate clusters at core density 50,000 and 35.}
\label{core}
\end{figure*}

\begin{figure}
\centering
\includegraphics[width=7.0cm, angle=0]{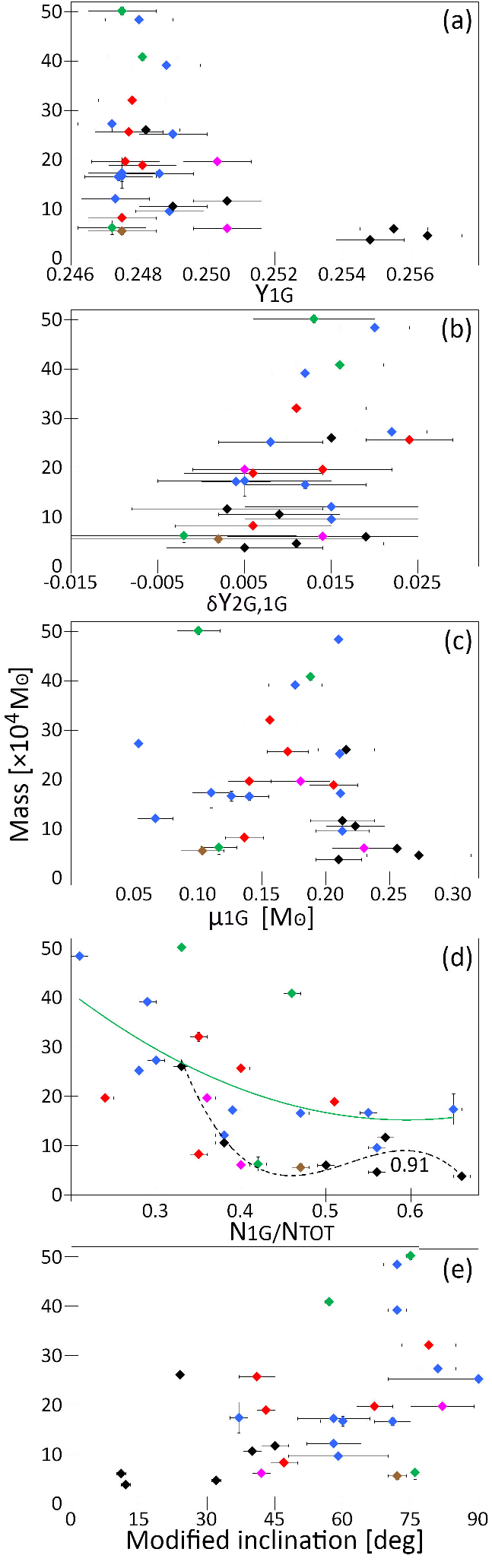}
\caption{The same as Fig.~\ref{feh} but for cluster mass vs some parameters.
}
\label{mass}
\end{figure}

\begin{figure*}
   \centering
   \includegraphics[width=14.5cm, angle=0]{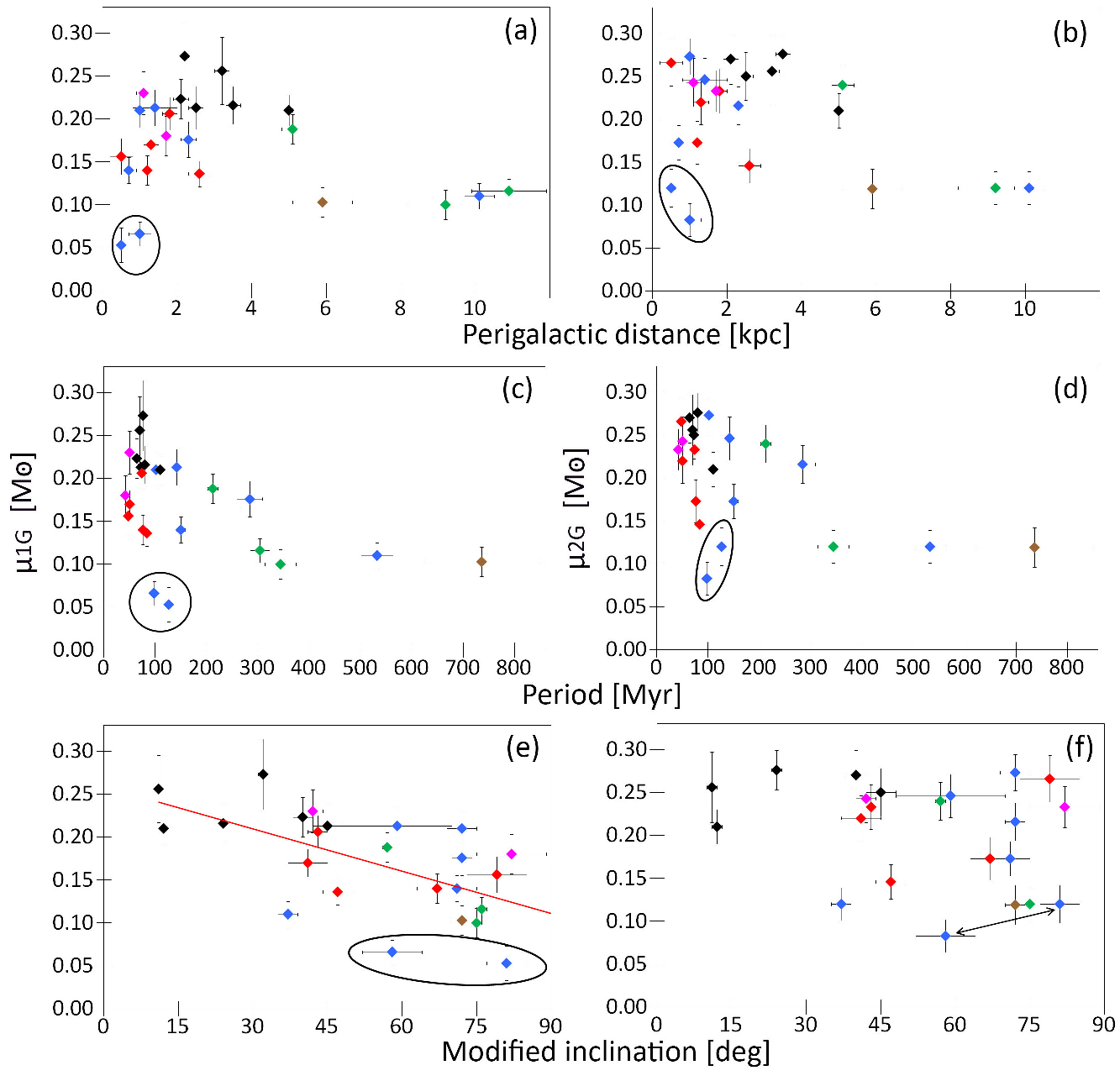}
   \caption{The same as Fig.~\ref{feh} but for $\mu_\mathrm{1G}$ (left column) and $\mu_\mathrm{2G}$ (right column) vs the orbital parameters of the clusters.
NGC\,6341 and NGC\,7099 are marked either by the black oval or by the arrow.}
\label{ml_orbit_}
\end{figure*}

Figs~\ref{feh}, \ref{y_ml}, \ref{core}, \ref{mass}, and \ref{ml_orbit_} present the most noticeable relations of the cluster parameters.
Their polynomial approximations are shown by different colours for in-situ, accreted, and all the clusters.
Their coefficients of determination $>0.5$ (i.e. for rather significant trends) are marked near the polynomial curves by the same colour.
Polynomial approximations with a lower coefficient of determination are shown in our figures in some cases in order to pay attention to interesting but less significant relations.
It is seen that, typically, all the three polynomial approximations for in-situ, accreted, and all the clusters follow each other.
Some figures with less noticeable relations are presented in Appendix~\ref{more}.

It is worth noting that some relations cannot be presented directly by a polynomial but they are seen as a non-uniform distribution of clusters in a plane of a pair of parameters. For example, clusters with a very high or very low core density indicate a low metallicity [$[$Fe$/$H$]<-1.4$, Fig.~\ref{core}~(a)], low $Y_\mathrm{1G}<0.249$ [Fig.~\ref{core}~(b)], as well as an old age $>12$ Gyr [Fig.~\ref{core}~(c)]. Among them, those with a very high/low core density have a high/low $\delta Y_\mathrm{2G,1G}$, peri- and apogalactic distance, respectively [Fig.~\ref{core}~(d)--(f)] (and, hence, orbital period due to a strong correlation between apogalactic distance and period).
Probably, this means that the transformation of a cluster into a loose or core-collapse state needs much time, prevents metal and helium enrichment of its 1G stars, and depends on the position of cluster orbit in the Galaxy.

Also, without direct relations, Fig.~\ref{mass} shows that the most massive clusters are exclusively accreted ones, with a relatively low percentage of 1G stars, steep orbits (i.e. with a high modified inclination), relatively high mass loss $\mu_\mathrm{1G}$, low $Y_\mathrm{1G}$, and high $\delta Y_\mathrm{2G,1G}$.
This suggests helium enrichment of the 1G and 2G stars in different processes. This is confirmed by the facts that clusters with a high core density (supposed core-collapse ones) have the lowest and highest helium enrichment of the 1G and 2G stars, respectively [Fig.~\ref{core}~(b) and (d)] and, in contrast to $Y_\mathrm{1G}$ in Fig.~\ref{y_ml}~(a), $\delta Y_\mathrm{2G,1G}$ is independent of mass loss in Fig.~\ref{y_ml}~(b). Note that the relation between $Y_\mathrm{1G}$ and mass loss in Fig.~\ref{y_ml}~(a) is induced by the relations between $[$Fe$/$H$]$ and $Y_\mathrm{1G}$ in Fig.~\ref{feh}~(a) and mass loss in Fig.~\ref{feh}~(c).

Direct relations between $[$Fe$/$H$]$ and $Y_\mathrm{1G}$ from models [as that from BaSTI in Fig.~\ref{feh}~(a)] seem to be realistic.
This means that the 1G stars follow the natural patterns of chemical evolution in the Galaxy, while the 2G stars do not, as follows from Fig.~\ref{feh}~(b) with a weak, if any, relation. 
Fig.~\ref{feh}~(b) and (e) and Fig.~\ref{mass}~(b) and (c) demonstrate a similar pattern of relations between $[$Fe$/$H$]$, cluster mass, $\delta Y_\mathrm{2G,1G}$, and mass loss $\mu_\mathrm{1G}$: most accreted clusters with medium $[$Fe$/$H$]$ and medium $\mu_\mathrm{1G}$ are distinguished by their high mass and high $\delta Y_\mathrm{2G,1G}$.
This suggests that the 2G stars were born and enriched by helium predominantly in rather dense environments or in massive clusters.
The importance of cluster mass is confirmed by a lower percentage of the 1G and, hence, higher percentage of the 2G stars in the massive clusters seen in Fig.~\ref{mass}~(d).

Variations of the mass loss with $[$Fe$/$H$]$ in Fig.~\ref{feh}~(c) and (d) are strong. These diagrams are similar to figure 5 from \citetalias{tailo2020}.
Although the non-linearity of these variations is attributed by \citetalias{tailo2020} to a difference between M3- and M13-like clusters, this can also be explained by a difference between in-situ (black and magenta) and accreted (other colours) clusters.
Namely, it seems that the accreted clusters demonstrate a significant increase of mass loss with $[$Fe$/$H$]$ in Fig.~\ref{feh}~(c) and (d) and, hence, with $Y_\mathrm{1G}$ in Fig.~\ref{y_ml}~(a), as well as with cluster mass in Fig.~ \ref{mass}~(c), while the in-situ clusters have rather high and nearly constant mass loss independent of metallicity.

Fig.~\ref{ml_orbit_} shows mass loss in dependence on some orbital parameters.
NGC\,6341 and NGC\,7099 (encircled or marked by the arrow) seem to be outliers relative to other clusters (this is also true for other orbital parameters).
Very low mass loss for NGC\,6341 and NGC\,7099 may be explained by their very low $[$Fe$/$H$]$ in combination with their core-collapse state in contrast to NGC\,5053 and NGC\,5466 with similar very low $[$Fe$/$H$]$ but much higher mass loss probably due to their loose state. 
However, another possible explanation is the following.
The HB stars of these clusters have a non-Gaussian distribution on mass (i.e. along the best-fitting BaSTI or any other ZAHB isochrone predicting mass) with an abrupt cutoff on the HB blue side. Such distribution is an indicator of the loss of low-mass HB stars. Such a loss is not taken into account by \citetalias{tailo2020} in their calculation of the mass loss, which, hence, may be underestimated. Accordingly, if the mass loss in NGC\,6341 and NGC\,7099 is higher, they would be among other clusters in Fig.~\ref{ml_orbit_}.

\begin{figure*}
\centering
\includegraphics[width=12.0cm, angle=0]{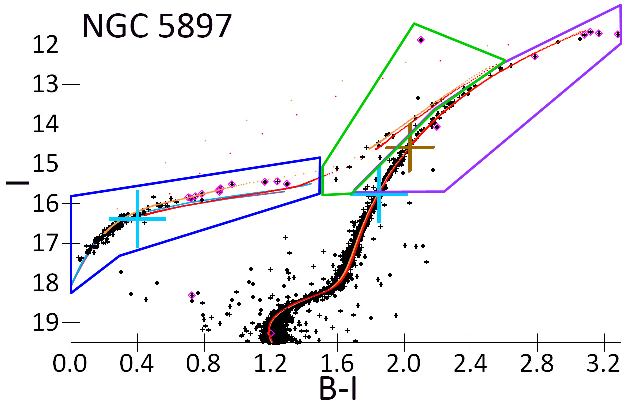}
\caption{A part of the $B-I$ vs $I$ CMD for the {\it Gaia} DR3 members from the \citetalias{stetson2019} data set for NGC\,5897 (black symbols).
The isochrones for $Y\approx0.25$ from BaSTI (red) and BaSTI ZAHB (purple), isochrones for $Y=0.275$ from BaSTI (orange) and BaSTI ZAHB (blue) are calculated with the best-fitting parameters from Table~\ref{cmds}.
Variable stars are shown by the magenta diamonds.
The left and right blue crosses mark the reference point on the ZAHB and the point on the RGB with the same luminosity, respectively.
The brown cross marks the RGB bump.
The blue, green, and purple polygons outline the CMD areas where all stars are assigned to the HB, AGB, and RGB, respectively.
}
\label{ngc5897_stetson_}
\end{figure*}

\begin{figure*}
\centering
\includegraphics[width=14.5cm, angle=0]{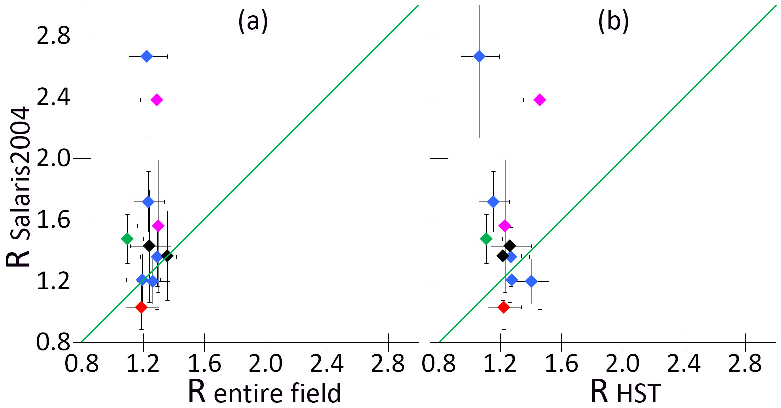}
\caption{$R$-parameter estimates from \citet{salaris2004} vs our estimates of $R$-parameter for eleven common clusters in
(a) the entire cluster field and (b) the part of the field covered by {\it HST}.
The green line represents the one-to-one correspondence.
Clusters of different association are marked by different colours (see the text).
}
\label{rr}
\end{figure*}

\section{Statistics of giants}
\label{giants}

\begin{table*}
\def\baselinestretch{1}\normalsize\footnotesize
\caption{The count of the RGB, HB, and AGB giants and $R$-parameter based on the {\it HST} or {\it Gaia} data sets, or in the entire cluster fields.
Only uncertainties for $R$ in the entire field are provided.
}
\label{rgbhbagb}
\[
\begin{tabular}{lcccccccccccc}
\hline
\noalign{\smallskip}
Cluster   & \multicolumn{3}{c}{$N_\mathrm{RGB}$} & \multicolumn{3}{c}{$N_\mathrm{HB}$} & \multicolumn{3}{c}{$N_\mathrm{AGB}$} & \multicolumn{3}{c}{$R$} \\
          & {\it HST} & {\it Gaia} & Total & {\it HST} & {\it Gaia} & Total & {\it HST} & {\it Gaia} & Total & {\it HST} & {\it Gaia} & Total \\
\hline
\noalign{\smallskip}
NGC\,288  &  41 &  59 & 100 &    50  &  74 & 124 &      9 &  11 &  20 & 1.22 & 1.25 & $1.240^{+0.108}_{-0.099}$ \\
\noalign{\smallskip}
NGC\,362  & 231 &  67 & 298 &    293 &  92 & 385 &    117 &  20 & 137 & 1.27 & 1.37 & $1.292^{+0.122}_{-0.110}$ \\
\noalign{\smallskip}
NGC\,1261 & 163 &  53 & 216 &    208 &  50 & 258 &     55 &  23 &  78 & 1.28 & 0.94 & $1.194^{+0.116}_{-0.105}$ \\
\noalign{\smallskip}
NGC\,5024 & 338 & 158 & 496 &    375 & 168 & 543 &     90 &  47 & 137 & 1.11 & 1.06 & $1.095^{+0.104}_{-0.095}$ \\
\noalign{\smallskip}
NGC\,5053 &  23 &  34 &  57 &     19 &  32 &  51 &      5 &   8 &  13 & 0.83 & 0.94 & $0.895^{+0.166}_{-0.142}$ \\
\noalign{\smallskip}
NGC\,5272 & 314 & 181 & 495 &    389 & 217 & 606 &     77 &  40 & 117 & 1.24 & 1.20 & $1.224^{+0.104}_{-0.095}$ \\
\noalign{\smallskip}
NGC\,5466 &  30 &  46 &  76 &     40 &  51 &  91 &      7 &  15 &  22 & 1.33 & 1.11 & $1.197^{+0.246}_{-0.201}$ \\
\noalign{\smallskip}
NGC\,5897 &  62 & 105 & 167 &     86 &  93 & 179 &     12 &  24 &  36 & 1.39 & 0.89 & $1.072^{+0.149}_{-0.130}$ \\
\noalign{\smallskip}
NGC\,5904 & 218 & 189 & 407 &    306 & 207 & 513 &     62 &  51 & 113 & 1.40 & 1.10 & $1.260^{+0.113}_{-0.103}$ \\
\noalign{\smallskip}
NGC\,6093 & 243 &  46 & 289 &    297 &  46 & 343 &     61 &  11 &  72 & 1.22 & 1.00 & $1.187^{+0.113}_{-0.103}$ \\
\noalign{\smallskip}
NGC\,6101 &  70 &  87 & 157 &     81 & 109 & 190 &     31 &  18 &  49 & 1.16 & 1.25 & $1.210^{+0.119}_{-0.108}$ \\
\noalign{\smallskip}
NGC\,6171 &  52 &  29 &  81 &     64 &  41 & 105 &     22 &  16 &  38 & 1.23 & 1.41 & $1.296^{+0.156}_{-0.137}$ \\
\noalign{\smallskip}
NGC\,6205 & 290 & 169 & 459 &    335 & 231 & 566 &     47 &  17 &  64 & 1.16 & 1.37 & $1.233^{+0.105}_{-0.096}$ \\
\noalign{\smallskip}
NGC\,6218 &  60 &  66 & 126 &     73 &  98 & 171 &     14 &  10 &  24 & 1.22 & 1.48 & $1.357^{+0.120}_{-0.109}$ \\
\noalign{\smallskip}
NGC\,6254 & 121 & 113 & 234 &    163 &  98 & 261 &     31 &  18 &  49 & 1.35 & 0.87 & $1.115^{+0.088}_{-0.081}$ \\
\noalign{\smallskip}
NGC\,6341 & 229 &  76 & 305 &    228 &  87 & 315 &     60 &   8 &  68 & 1.00 & 1.14 & $1.033^{+0.107}_{-0.096}$ \\
\noalign{\smallskip}
NGC\,6352 &  32 &  40 &  72 &     53 &  34 &  87 &     14 &  17 &  31 & 1.66 & 0.85 & $1.208^{+0.332}_{-0.255}$ \\
\noalign{\smallskip}
NGC\,6362 &  65 &  56 & 121 &     82 &  68 & 150 &     17 &  13 &  30 & 1.26 & 1.21 & $1.240^{+0.141}_{-0.125}$ \\
\noalign{\smallskip}
NGC\,6366 &   9 &  41 &  50 &     22 &  42 &  64 &     10 &  13 &  23 & 2.44 & 1.02 & $1.280^{+0.385}_{-0.288}$ \\
\noalign{\smallskip}
NGC\,6397 &  30 &  80 & 110 &     41 &  80 & 121 &      5 &  14 &  19 & 1.37 & 1.00 & $1.100^{+0.322}_{-0.247}$ \\
\noalign{\smallskip}
NGC\,6541 & 195 & 120 & 315 &    263 &  88 & 351 &     48 &  16 &  64 & 1.35 & 0.73 & $1.114^{+0.110}_{-0.100}$ \\
\noalign{\smallskip}
NGC\,6723 & 118 &  66 & 184 &    172 &  65 & 237 &     42 &  26 &  68 & 1.46 & 0.98 & $1.288^{+0.119}_{-0.107}$ \\
\noalign{\smallskip}
NGC\,6752 & 116 & 118 & 234 &    160 & 151 & 311 &     17 &  17 &  34 & 1.38 & 1.28 & $1.329^{+0.124}_{-0.112}$ \\
\noalign{\smallskip}
NGC\,6779 & 118 &  57 & 175 &    162 &  63 & 225 &     54 &  10 &  64 & 1.37 & 1.11 & $1.286^{+0.145}_{-0.128}$ \\
\noalign{\smallskip}
NGC\,6809 &  53 & 147 & 200 &     63 & 154 & 217 &     14 &  26 &  40 & 1.19 & 1.05 & $1.085^{+0.123}_{-0.110}$ \\
\noalign{\smallskip}
NGC\,6838 &  20 &  25 &  45 &     26 &  36 &  62 &     13 &  19 &  32 & 1.30 & 1.44 & $1.378^{+0.260}_{-0.213}$ \\
\noalign{\smallskip}
NGC\,7099 &  98 &  36 & 134 &    104 &  60 & 164 &     10 &  13 &  23 & 1.06 & 1.67 & $1.224^{+0.133}_{-0.119}$ \\
\hline
\end{tabular}
\]
\end{table*}

We deliberately chose clusters, whose RGB, HB, and AGB stars are rather confidently separated in any CMD under consideration (this is not the case, for example, for NGC\,2808, which, hence, is not considered). In particular, the HB and AGB are separated by the end of the core helium burning at the corresponding theoretical evolution track.
As a result, the position of each giant in each CMD w.r.t. the best-fitting BaSTI ZAHB, AGB, and RGB, even taking into account its photometric uncertainty, allows us to classify this giant as an HB, AGB, or RGB star with a rather high probability.
An example of such classification for the {\it Gaia} members in the \citetalias{stetson2019} data set for NGC\,5897 is presented in Fig.~\ref{ngc5897_stetson_}.
It is seen that only a few giants can be misclassified for this cluster. This is similar for other CMDs and other clusters.

This allows us to count the HB and AGB stars, as well as the RGB stars brighter than the ZAHB.
To cut the RGB at the ZAHB level escaping the uncertainties of bolometric corrections and colour--$T_\mathrm{eff}$ relations, we cut it at the luminosity (but not the magnitude) of the best-fitting BaSTI ZAHB. This cut level is shown in Fig.~\ref{ngc5897_stetson_} by the blue cross on the RGB corresponding to the blue cross on the ZAHB.
To cover the entire cluster field and count each giant once, we cross-identify the data sets to each other and eliminate duplicates.

It is worth noting that, since the ZAHB luminosity depends on mass loss, which varies considerably for the clusters (see Table~\ref{mix}), we use the BaSTI ZAHB with a stochastic mass loss to derive the ZAHB luminosity for cluster's $[$Fe$/$H$]$, age, $Y_\mathrm{mix}$, and $\mu_\mathrm{mix}$.
For NGC\,362, NGC\,1261, and NGC\,5897 without mass loss estimates from \citetalias{tailo2020}, we adopt $\mu_\mathrm{mix}=0.24$, 0.24, and 0.14 solar masses, respectively, calculated with the polynomials from Fig.~\ref{feh}~(c) and (d) as functions of their $[$Fe$/$H$]$.

Table~\ref{rgbhbagb} presents the numbers $N_\mathrm{RGB}$, $N_\mathrm{HB}$, and $N_\mathrm{AGB}$ of the RGB, HB, and AGB stars, respectively,\footnote{The lists of these giants are
available upon request.}
for (i) the {\it HST} data sets covering the central parts of the cluster fields within about 2 arcmin from the centres, (ii) the {\it Gaia} data sets (also taking into account the {\it Gaia} members in the \citetalias{stetson2019} data sets) minus stars common to the {\it HST} and {\it Gaia} data sets, which, hence, cover, on the contrary, the remaining outer parts of the fields, and (iii) the entire fields.
The RGB stars include only those with luminosity higher than the ZAHB.
Also, Table~\ref{rgbhbagb} provides the $R$-parameter for the inner, outer parts, and the entire fields.

To calculate the uncertainty of $R$, we take into account the photometric uncertainty, DR uncertainty, uncertainty of the derived parameters of the best-fitting isochrone, uncertainty of the classification (RGB, HB, or AGB), probable contamination by remaining non-members (this appears important only for NGC\,362), membership probability (we assign a probability of $1$ to the {\it Gaia} members and {\it HST} stars with probability $>90\%$, while a probability of $0.5$ to the {\it HST} stars with undefined membership), underestimation of the {\it HST} RGB count due to saturation of the RGB tip (this appears important only for the nearest metal-poor clusters NGC\,6397 and NGC\,6752), and uncertainties of the adopted $Y_\mathrm{mix}$ and $\mu_\mathrm{mix}$ (the latter makes a significant contribution to the uncertainty of $R$ for all the clusters).

Another source of uncertainty is due to the RGB bump, an overdensity of the RGB stars in any CMD arising from their temporary decrease in luminosity during its overall increase \citep{salaris2004}.
The bump is predicted by BaSTI and other models at a luminosity brighter or fainter than that of the ZAHB for a low or high $[$Fe$/$H$]$, respectively.
Consequently, this can provide systematically higher or lower $N_\mathrm{RGB}$ and, hence, lower or higher $R$ for metal-poor and metal-rich clusters, respectively \citep{salaris2004}.
For example, Fig.~\ref{ngc5897_stetson_} shows the bump (brown cross) for rather metal-poor NGC\,5897 brighter than the ZAHB level on the RGB (blue cross).
The best-fitting BaSTI isochrones for the clusters under consideration show that the RGB bump coincides with the ZAHB luminosity level for $[$Fe$/$H$]\approx0.8$, meaning that NGC\,6171, NGC\,6362 and NGC\,6723 are slightly metal-poorer, while NGC\,6352, NGC\,6366, and NGC\,6838 are slightly metal-richer than this metallicity boundary.
We calculate that the RGB bump affects the dependence of $R$ on $[$Fe$/$H$]$ within the total uncertainty of our $R$ estimates presented in Table~\ref{rgbhbagb}, since only few, if any, RGB stars appear at the bump for the clusters under consideration.

The last determination of homogeneous $R$ for many clusters is presented by \citet{salaris2004}. It is based on observations with the {\it HST} Wide Field and Planetary Camera 2 (WFPC2).
Fig.~\ref{rr} compares our $R$ estimates with those from \citet{salaris2004} for eleven common clusters in (a) the entire cluster field and (b) the part of the field covered by {\it HST}.
The latter may be more appropriate since the {\it HST} WFPC2 observations used by \citet{salaris2004} cover nearly the same field within about 2 arcmin from the cluster centres as the {\it HST} ACS observations used by us do. However, Fig.~\ref{rr} indicates an agreement of our and \citet{salaris2004} $R$ estimates within their $1\sigma$ uncertainties for six and five among eleven common clusters for the entire and {\it HST} fields, respectively. It is worth noting that the uncertainty of the $R$ estimates from \citet{salaris2004} is typically much larger than ours. This balance of the uncertainties seems to be realistic, since the ACS observations are much more accurate than those from the WFPC2, our data are much better cleaned of non-members, and models used are more advanced.

\begin{figure*}
\centering
\includegraphics[width=14.5cm, angle=0]{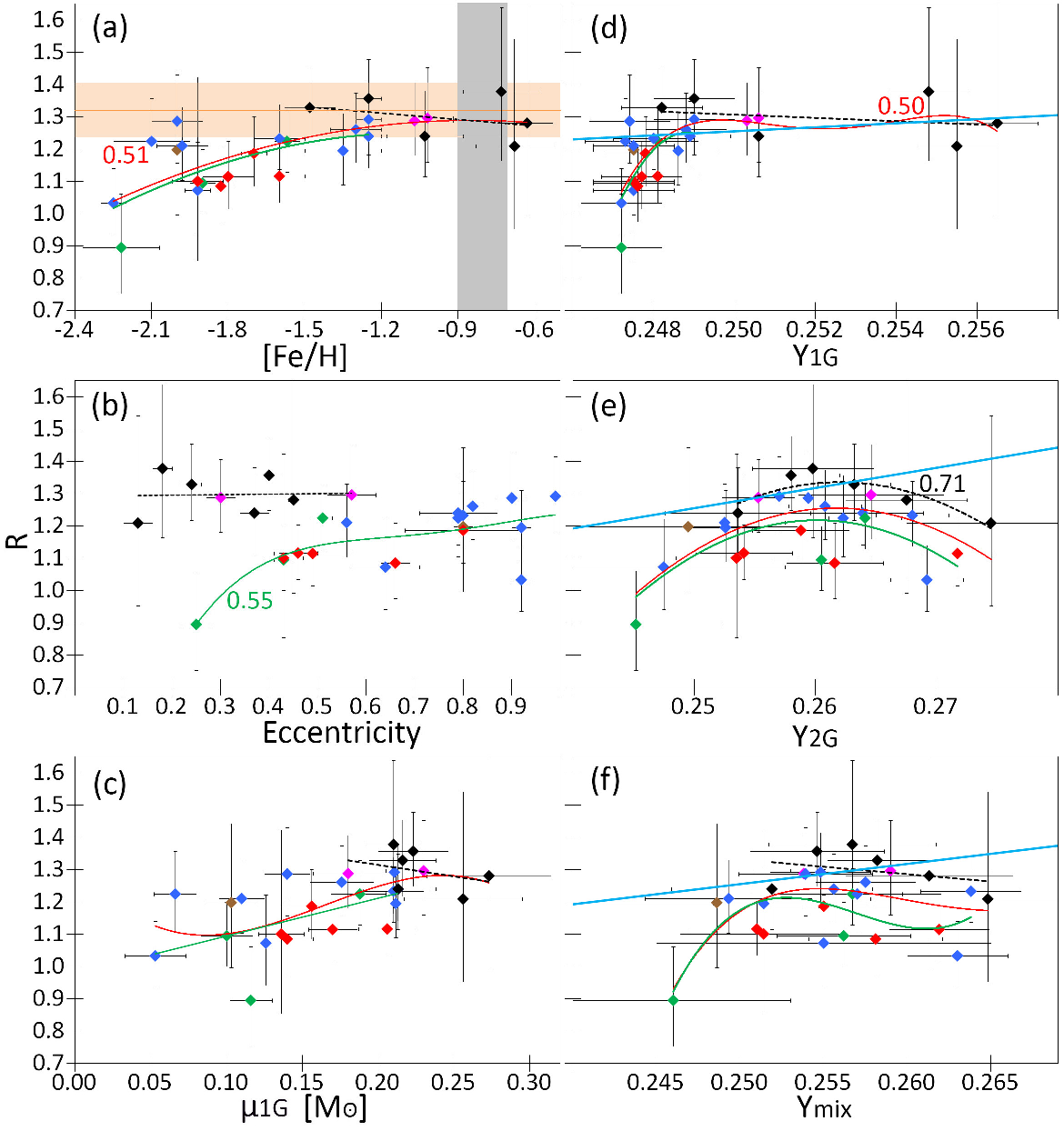}
\caption{$R$-parameter as functions of some cluster parameters.
Clusters of different association are marked by different colours (see the text).
The orange line and error band show the constant $R=1.32$ with its uncertainty with which the estimates for all in-situ and most accreted clusters agree.
The black dotted, green solid, and red solid curves show polynomial approximations for the in-situ, accreted, and all clusters, respectively.
The coefficients of determination $>0.5$ are marked near the polynomial curves by the same colour.
The grey area shows the $-0.9<[$Fe$/$H$]<-0.7$ range where the RGB bump may change $R$.
The light blue line shows the $Y$ vs $R$ relation from Equation (1) of \citet{ayala2014} with $g_{a\gamma}=0.68\times10^{-10}$ GeV$^{-1}$.
}
\label{r}
\end{figure*}

For many years, it has been thought that $R$ depends mainly on $Y$, since, at a given metallicity, the higher the $Y$, the brighter the ZAHB, the smaller $N_\mathrm{RGB}$ brighter than the ZAHB and, hence, the larger the value of $R$ \citep{salaris2004}. Furthermore, \citet{ayala2014} (see also \citealt{carenza2025}) note that theoretically $R$ must be almost independent of cluster metallicity and age. However, the aforementioned direct relation between $Y_\mathrm{1G}$ and $[$Fe$/$H$]$ suggests a dependence of $R$ on the latter due to a dependence on the former.
Now we have much more accurate data and theoretical predictions to check in detail what parameters $R$ depends on.

Fig.~\ref{r} shows $R$ as functions of several parameters. Dependences on other parameters are less significant. It is seen that the $R$ estimates for all eight in-situ and most (12 among 19) accreted clusters agree within $1\sigma$ with a rather high constant value $R=1.32\pm0.09$ shown by the orange line with error band. 
The grey area in Fig.~\ref{r}~(a) denotes the $-0.9<[$Fe$/$H$]<-0.7$ range where the RGB bump may cause an increase in $R$. However, it is seen that, as our calculation predicts, any such increase is within the $R$ uncertainties.
Seven outliers with a lower $R$ are metal-poor ($[$Fe$/$H$]<-1.6$), helium-poor ($Y_\mathrm{1G}<0.2481$), and rather old ($>12.4$ Gyr) accreted clusters with a medium orbital eccentricity. These results can be interpreted as a constant $R=1.31^{+0.06}_{-0.05}$ for all the in-situ clusters and a trend for the accreted ones due to a homogeneous environment in the Galaxy in contrast to heterogeneous origins and environments of the accreted clusters. To establish the argument for this trend, we need more cluster statistics and/or estimates of higher accuracy.

Figs~\ref{r}~(d) and (f) show that our $R$ estimates for most clusters, except several helium-poor ones, agree with the theoretical trend in Equation (1) of \citet{ayala2014} for $R$ vs $Y$, if anomalous energy losses in the HB stars correspond to the axion-photon coupling $g_{a\gamma} \sim 0.68 \times10^{-10}\,\mathrm{GeV}^{-1}$.

For comparison with previous work, we also derive a cleaner estimate following the approach of \citet{ayala2014}. We select eight in-situ clusters and apply the procedure described by \citet{troitsky2025}, obtaining $g_{a\gamma}=(0.66^{+0.11}_{-0.13})\times10^{-10}\,\mathrm{GeV}^{-1}$. This value is consistent with earlier estimates \citep{ayala2014,dolan2022,troitsky2025}. However, the results of the present study suggest that this standard application of the $R$-parameter method to globular clusters is incomplete and should be modified.

Rather large number of clusters under consideration allows one to see that our Fig.~\ref{r}~(a) mainly reproduces the \citet{salaris2004}'s dependence of $R$ on $[$Fe$/$H$]$ shown in their Figure 1 by individual estimates but not by curves, which are seem to be too smooth (also, see Fig.~\ref{rr}).
This is true for both the metallicity scales from \citet{zinn1984} and \citet{carretta1997} used by \citet{salaris2004} in their Figure 1.

Since (i) our results generally reproduce the $R$ vs $[$Fe$/$H$]$ relation from \citet{salaris2004}, (ii) there is the direct relation between $[$Fe$/$H$]$ and $Y_\mathrm{1G}$, and (iii) the \citet{ayala2014} results are based on the \citet{salaris2004} data, then the $R$ vs $Y_\mathrm{1G}$ or $R$ vs $Y_\mathrm{mix}$ relations seen in Figs~\ref{r}~(d) and (f), respectively, are expected to be present both in the \citet{salaris2004} and \citet{ayala2014} results.
Furthermore, Fig.~\ref{r} shows that among the accreted clusters, the low-energy (red symbols) and Helmi stream (green symbols) clusters tend to provide lower estimates (average $R=1.12$ and 1.07, respectively) than those for the Gaia-Sausage-Enceladus (blue symbols) and Sequoia (brown symbol) clusters (average $R=1.20$ for both). This suggests that progenitor properties may affect cluster's $R$. Hence, Equation (1) from \citet{ayala2014} may be just a rough incomplete approximation of a real $R$ vs $Y$ relation decreasing for helium-poor clusters.

\begin{figure*}
\centering
\includegraphics[width=14.5cm, angle=0]{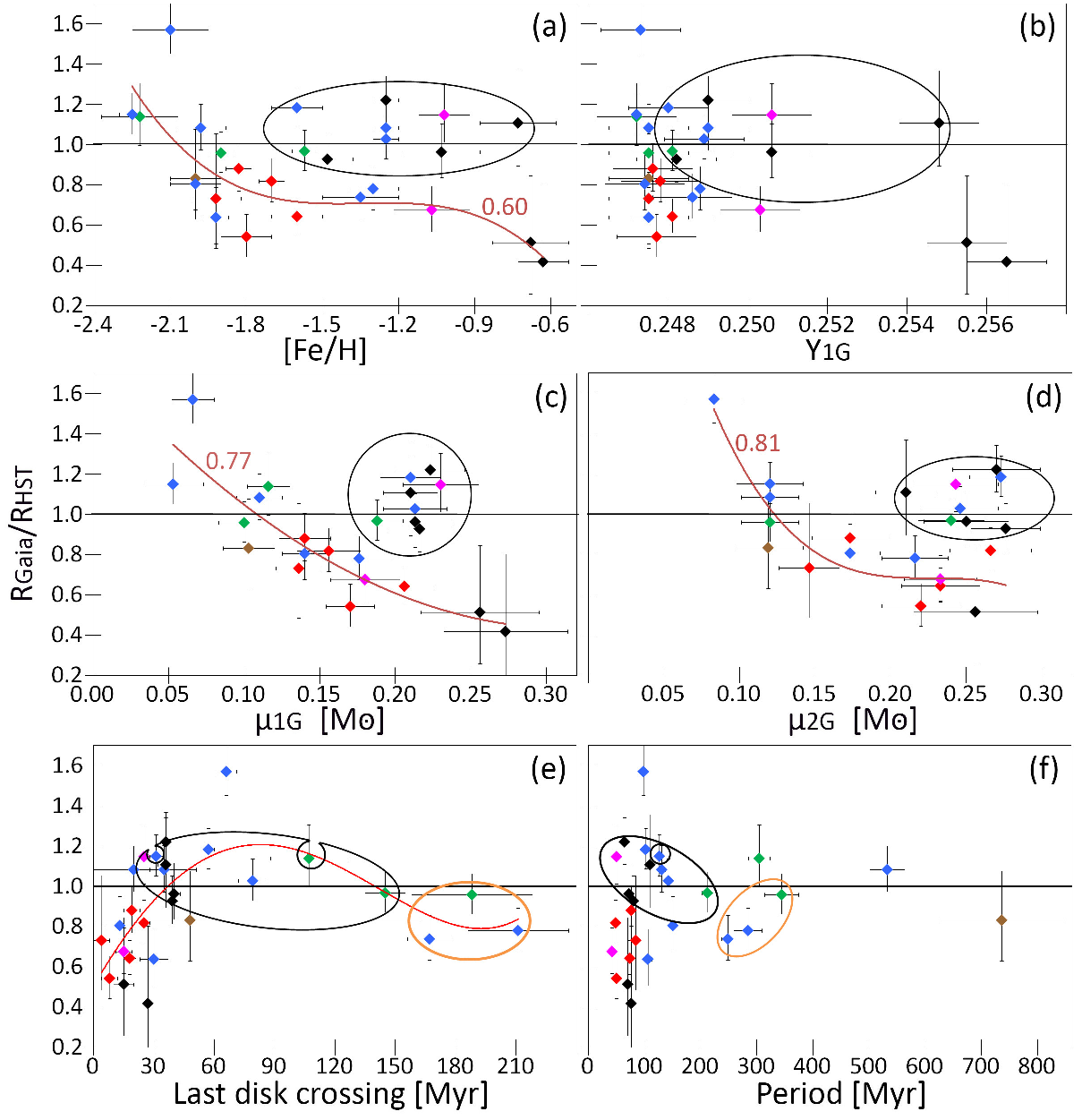}
\caption{$R_{Gaia}/R_{HST}$ as functions of some cluster parameters.
Clusters of different association are marked by different colours (see the text).
The solid black line shows the level $R_{Gaia}/R_{HST}=1$.
The black and orange circles encircle the eight special clusters and three clusters with the oldest last disk crossing, respectively (see the text).
The red and brown curves show polynomial approximations for all the clusters and those except the eight special clusters, respectively.
The coefficients of determination $>0.5$ are marked near the polynomial curves by the same colour.
}
\label{gaia_hst_ratio}
\end{figure*}

\subsection{$R_{Gaia}/R_{HST}$ ratio}
\label{ratio}

It is interesting to consider the ratio $R_{Gaia}/R_{HST}$ based on the data from Table~\ref{rgbhbagb}. Since the {\it HST} and {\it Gaia} data cover inner 2 arcmin within the cluster centres and remaining outer cluster field, respectively, the ratio $R_{Gaia}/R_{HST}$ may inform about a dependence of $R$ on the distance from the cluster centre.
Fig.~\ref{gaia_hst_ratio} shows dependences of $R_{Gaia}/R_{HST}$ on some cluster parameters.
A group of eight clusters (NGC\,288, NGC\,5272, NGC\,6205, NGC\,6218, NGC\,6362, NGC\,6752, NGC\,6838, and probably NGC\,362, which has no mass loss estimates from \citetalias{tailo2020}) separates from the remaining ones in some diagrams of Fig.~\ref{gaia_hst_ratio}. Accordingly, 19 remaining clusters show rather high coefficients of determination for the relations between $R_{Gaia}/R_{HST}$ and some parameters.

These relations in Fig.~\ref{gaia_hst_ratio}~(a)--(d) show that, generally, the higher the $[$Fe$/$H$]$, $Y_\mathrm{1G}$, $\mu_\mathrm{1G}$ or $\mu_\mathrm{2G}$, the more stars are lost from cluster periphery compared to its center.
Taking into account the correlations between $[$Fe$/$H$]$, $Y_\mathrm{1G}$, $\mu_\mathrm{1G}$, and $\mu_\mathrm{2G}$ from Fig.~\ref{feh}, all the relations in Fig.~\ref{gaia_hst_ratio}~(a)--(d) can be explained by one process: a higher mass loss at the RGB produces HB stars of a lower mass and, hence, a higher fraction of low-mass HB stars, which can be lost more effectively from cluster periphery compared to its center.

To understand the nature of the eight outliers, the dependence of $R$ on the time after the last disk crossing in Fig.~\ref{gaia_hst_ratio}~(e) is the most important. This suggests that any disk crossing eliminates more HB stars, as giants with the least mass, from the periphery of any cluster than from its centre, which results in a minimal $R_{Gaia}/R_{HST}$ just after the crossing. Then the HB at cluster periphery is restored due to evolution of the RGB stars into the HB ones. Accordingly, for a time equal to an average lifetime of stars on the HB, the income to the HB from the RGB is greater than the outcome from the HB to the AGB. Consequently, a rise of $R_{Gaia}/R_{HST}$ in Fig.~\ref{gaia_hst_ratio}~(e) allows us to estimate this time as about 60--80 Myr. The increase of $R_{Gaia}/R_{HST}$ from about 0.6 to about 1.2 during this time allows us to estimate that about half of the periphery HB stars are typically lost during the crossing. When the periphery HB is completely restored, $R_{Gaia}/R_{HST}$ becomes nearly constant, which is seen in Fig.~\ref{gaia_hst_ratio}~(e) for about 70--110 Myr after the crossing.
Thus, Fig.~\ref{gaia_hst_ratio}~(e) shows that the eight special clusters differ from the others by their recently restored HBs.
However, three rightmost clusters NGC\,1261, NGC\,5024, and NGC\,5904 in Fig.~\ref{gaia_hst_ratio}~(e), apparently having long-restored HBs, show a decrease of $R_{Gaia}/R_{HST}$. This may be explained by Fig.~\ref{gaia_hst_ratio}~(f) indicating that these three clusters have much longer orbital periods than those eight clusters. Probably, the longer the period, the more low-mass stars are lost from cluster periphery. On the other hand, Fig.~\ref{gaia_hst_ratio}~(f) shows the lowest $R_{Gaia}/R_{HST}$ ratio for clusters with the shortest periods. This may mean that more HB stars are eliminated from cluster periphery by not only a recent disk crossing but also by frequent crossings.

\section{Conclusions}
\label{conclusions}

We fitted CMDs based on the \citetalias{stetson2019}, {\it Hubble Space Telescope (HST)}, and {\it Gaia} DR3 data sets for members of 22 Galactic globular clusters by the BaSTI and DSED theoretical isochrones. This allowed us to derive metallicity $[$Fe$/$H$]$, age, distance from the Sun, and reddening $E(B-V)$ for these clusters. Combining these results with those for other five clusters from \citetalias{paper7}, we considered some cluster statistics and relations between their parameters.
We took into account the association of the in-situ clusters with the Galaxy and accreted ones with other progenitors.
Five processes appear most important for the relations between cluster parameters: chemical enrichment and mass loss, both different between the 1G and 2G stars,
and the loss of low-mass stars during cluster evolution.
We concluded that 1G follows the general patterns of chemical evolution in the Galaxy, while the 2G stars seem to be born from a helium-enriched medium
predominantly in rather dense environments or in massive clusters.
Mass loss in the accreted clusters significantly increases with $[$Fe$/$H$]$, $Y_\mathrm{1G}$, and cluster mass, while the in-situ clusters have nearly constant and rather high mass loss.

We counted the RGB, HB, and AGB giants in the centres, periphery, and entire fields of the clusters. This allowed us to calculate the ratio $R$ between the numbers of the HB and RGB stars with unprecedented accuracy and consider its relations with other parameters. These relations appear much stronger for accreted clusters rejecting a constant $R$ value for them, which may be due to their heterogeneous origin and environment. 
For in-situ clusters, the constant $R=1.31^{+0.06}_{-0.05}$ can be accepted, which would imply the observation of anomalous energy losses in the HB stars corresponding to the axion-photon coupling $g_{a\gamma}=(0.66^{+0.11}_{-0.13})\times10^{-10}\,\mathrm{GeV}^{-1}$.
However, as we discussed in Sect.~\ref{giants}, this relation has to be updated. Obtaining state-of-the-art constraints on $g_{a\gamma}$ will be the subject of our forthcoming work.
Our comparison of the $R$-parameter estimates in cluster centres and periphery suggests that the periphery HB is emptied by about half when cluster crosses the Galactic disk, then the HB restores for 60--80 Myr.
We intend to expand this study by more clusters for better statistics and detailed investigation of relations between cluster parameters.

\normalem
\begin{acknowledgements}
We acknowledge financial support from the Russian Science Foundation (grant no. 22--12--00253--P).

We thank the referee for a constructive and very useful report.
We thank
Armando Arellano Ferro and Nikolai Samus for very fruitful discussion of the cluster RR~Lyrae stars,
Anisa Bajkova and Vadim Bobylev for discussion of cluster orbital parameters,
Santi Cassisi for providing the valuable BaSTI isochrones with his exceptionally useful comments,
Aaron Dotter for his comments on DSED,
Maxim Khovritchev for technical support,
Alexey Kudashov, Konstantin Postnov, and Georg Raffelt for illuminating discussions of new-physics constraints from stellar evolution,
Peter Stetson for providing and having discussion of his valuable photometry,
Don VandenBerg for discussion of many aspects of globular clusters,
and Eugene Vasiliev for his very useful comments on the globular cluster properties.

This work has made use of BaSTI and DSED web tools;
Filtergraph \citep{filtergraph}, an online data visualization tool developed at Vanderbilt University through the Vanderbilt Initiative in
Data-intensive Astrophysics (VIDA) and the Frist Center for Autism and Innovation (FCAI, \url{https://filtergraph.com});
the resources of the Centre de Donn\'ees astronomiques de Strasbourg, Strasbourg, France (\url{http://cds.u-strasbg.fr}), including the SIMBAD database \citep{simbad},
the VizieR catalogue access tool \citep{vizier} and the X-Match service;
observations made with the NASA/ESA {\it Hubble Space Telescope};
data from the European Space Agency (ESA) mission {\it Gaia} (\url{https://www.cosmos.esa.int/gaia}), processed by the {\it Gaia} Data Processing and Analysis Consortium
(DPAC, \url{https://www.cosmos.esa.int/web/gaia/dpac/consortium}), and {\it Gaia} archive website (\url{https://archives.esac.esa.int/gaia}).

\end{acknowledgements}

\bibliographystyle{raa}
\bibliography{bibtex}

\clearpage
\appendix

\section{All CMDs under consideration}
\label{addcmds}

\begin{figure*}
\includegraphics[width=14.5cm, angle=0]{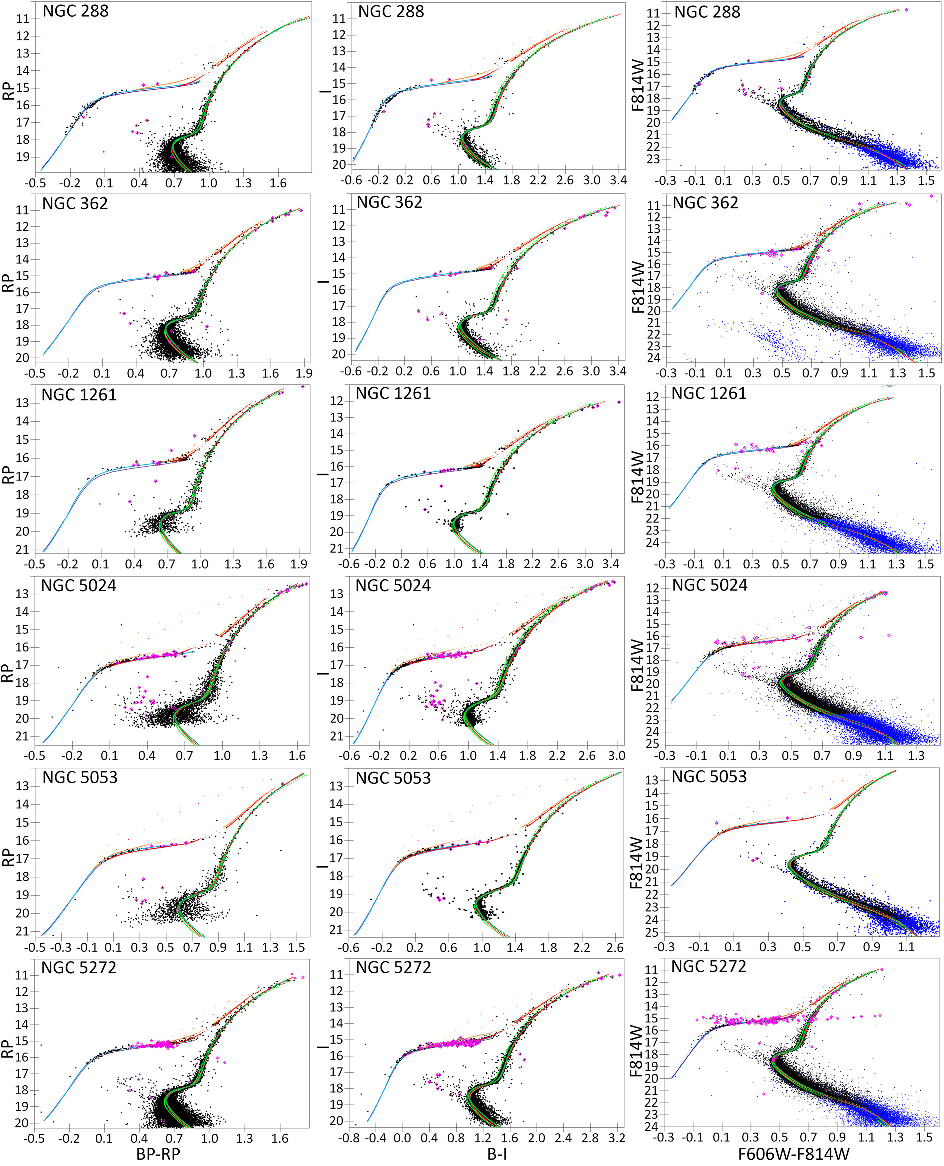}
\caption{The CMDs for the clusters. Left: $G_\mathrm{BP}-G_\mathrm{RP}$ vs $G_\mathrm{RP}$ from the {\it Gaia} DR3,
centre: $B-I$ vs $I$ from the \citetalias{stetson2019} data set cross-identified with the cluster members from {\it Gaia} DR3,
right: $F606W-F814W$ vs $F814W$ from \citetalias{nardiello2018} for cluster members with membership probability $>0.9$ (black symbols) and undefined membership probability $=-1$ (blue symbols).
The isochrones for $Y\approx0.25$ from BaSTI (red), BaSTI ZAHB (purple), and DSED (green), isochrones for $Y=0.275$ from BaSTI (orange) and BaSTI ZAHB (blue), as well as isochrones for $Y=0.33$ from DSED (luminous green) are calculated with the best-fitting parameters from Table~\ref{cmds}.
Variable stars are shown by the magenta diamonds.
}
\label{cmd1}
\end{figure*}

\begin{figure*}
\includegraphics[width=14.5cm, angle=0]{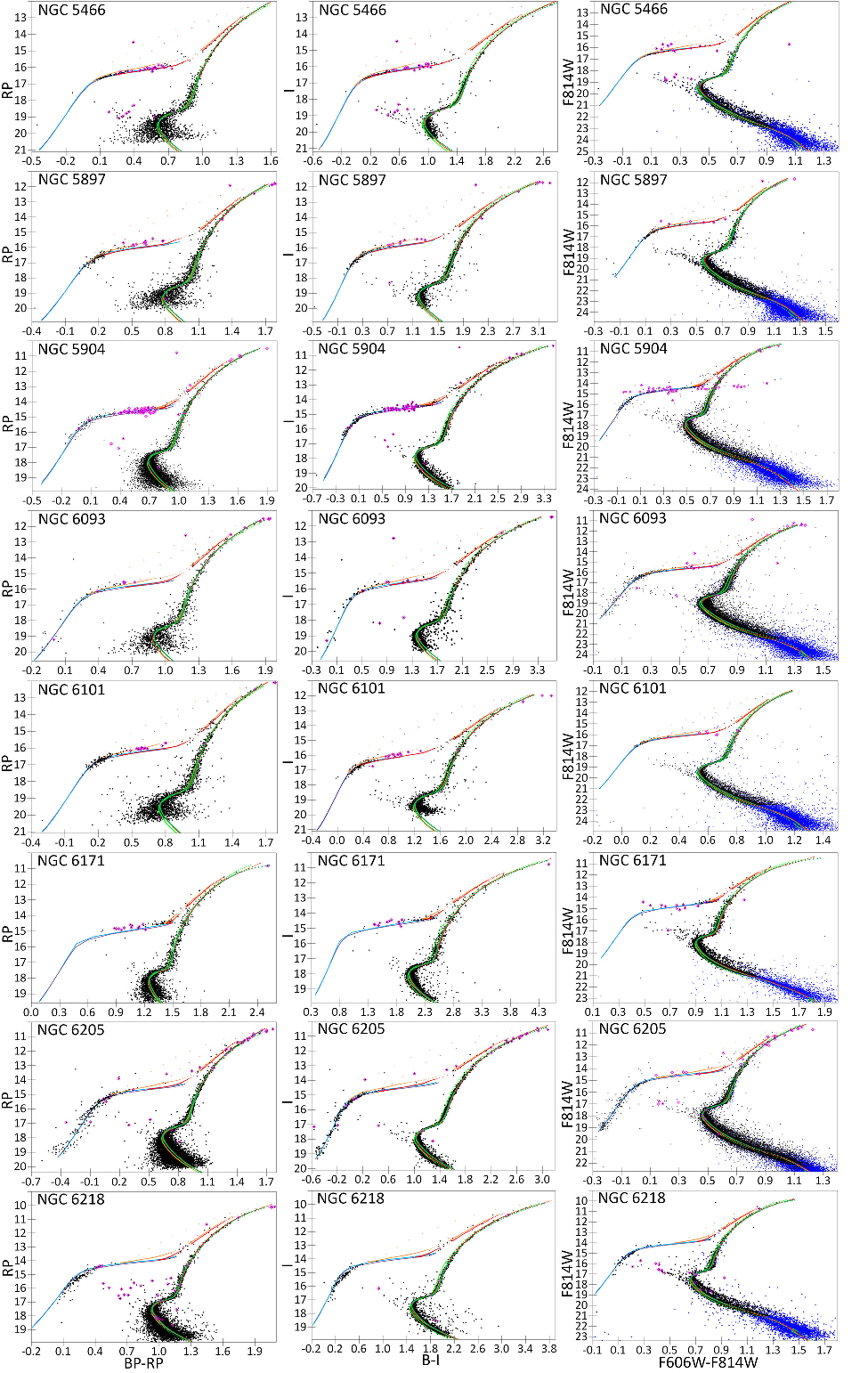}
\caption{The same as Fig.~\ref{cmd1} but for other clusters.
}
\label{cmd2}
\end{figure*}

\begin{figure*}
\includegraphics[width=14.5cm, angle=0]{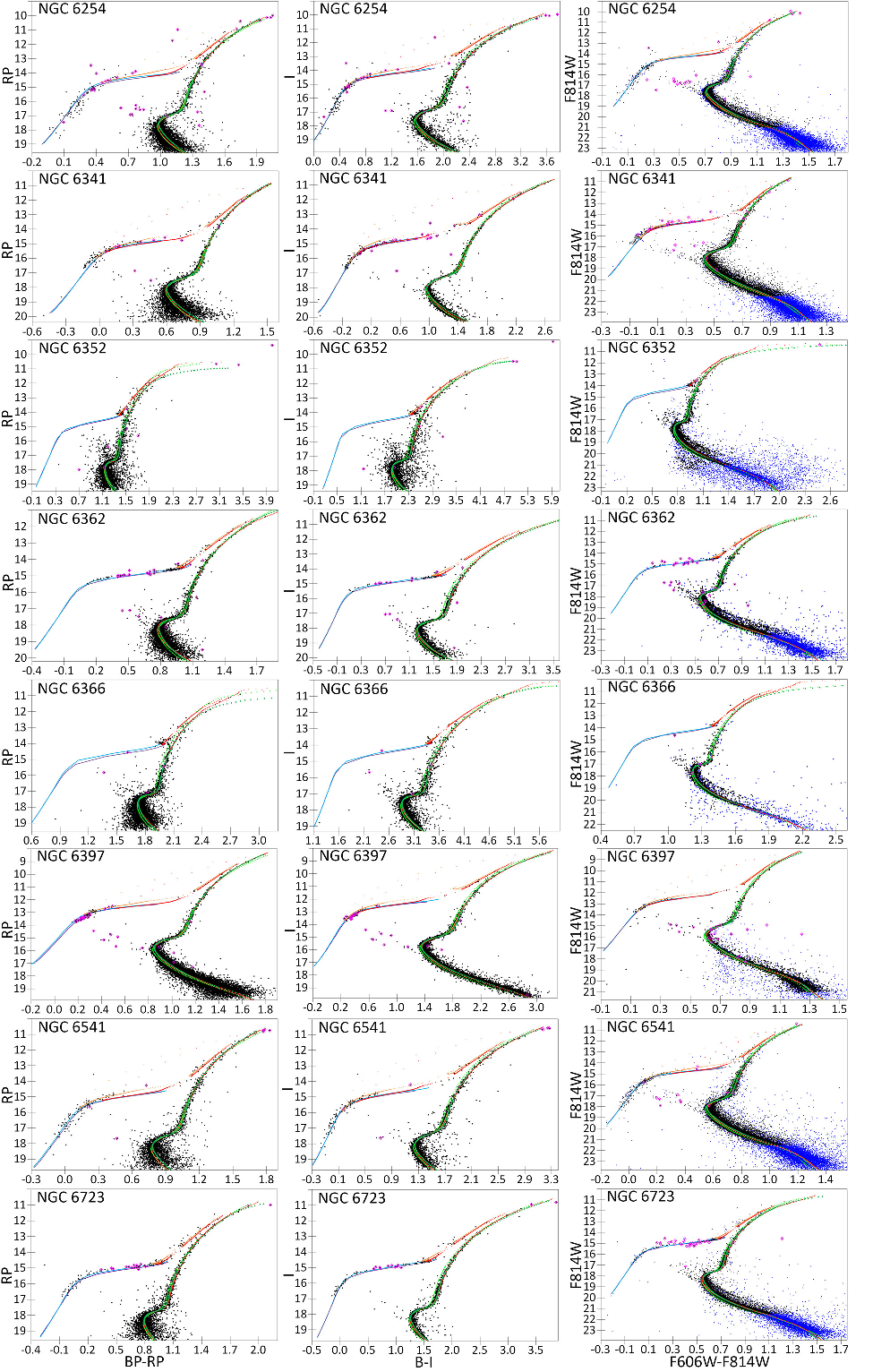}
\caption{The same as Fig.~\ref{cmd1} but for other clusters.
}
\label{cmd3}
\end{figure*}

\begin{figure*}
\includegraphics[width=14.5cm, angle=0]{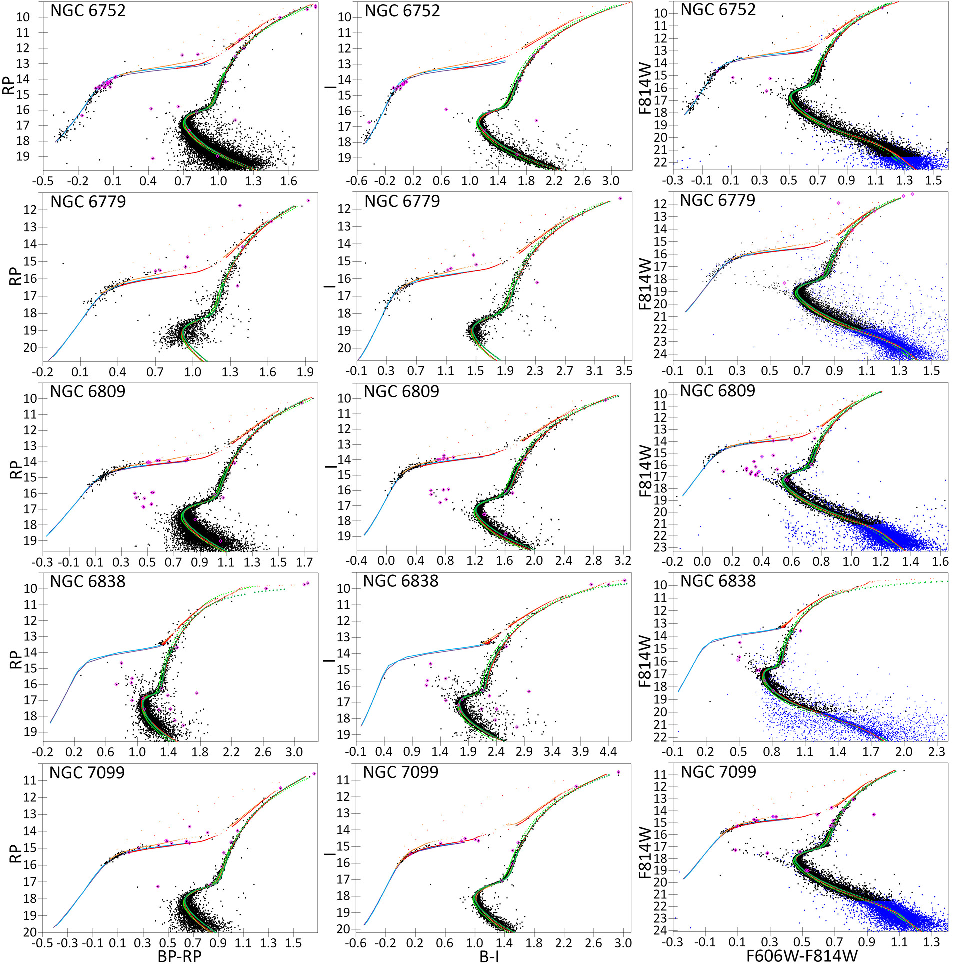}
\caption{The same as Fig.~\ref{cmd1} but for other clusters.
}
\label{cmd4}
\end{figure*}

We present all the CMDs with our best isochrone fitting in Figs~\ref{cmd1}, \ref{cmd2}, \ref{cmd3}, and \ref{cmd4}.

\section{The best-fitting parameters}
\label{best}

Table~\ref{cmds} presents the best solutions of our isochrone fitting for two models and all the CMDs.

Table B.1 The results of our isochrone fitting.
In all the CMDs, the observed colour is the abscissa and the observed magnitude in the redder filter is the ordinate.
Each derived reddening is converted into $E(B-V)$ using extinction coefficients from
\citet{casagrande2014,casagrande2018a,casagrande2018b} or \citet{ccm89}.
Age is in Gyr, cluster distance from the Sun $D$ is in kpc.
\begin{longtable}{lcccccccc}
\hline
\noalign{\smallskip}
\label{cmds}
   & \multicolumn{4}{c}{BaSTI} & \multicolumn{4}{c}{DSED} \\
Data set and colour                      & $[$Fe$/$H$]$ & Age  & $R$  & $E(B-V)$   & $[$Fe$/$H$]$ & Age  & $D$  & $E(B-V)$   \\
\hline
\noalign{\smallskip}
   & \multicolumn{8}{c}{NGC\,288} \\
\noalign{\smallskip}
\citetalias{nardiello2018} $F606W-F814W$ & $-1.2$ & 13.0 & 8.80 & $0.010$     & $-1.3$ & 13.0 & 8.80 & $0.029$ \\
{\it Gaia} $G_\mathrm{BP}-G_\mathrm{RP}$ & $-1.2$ & 13.5 & 8.75 & $0.011$     & $-1.3$ & 13.0 & 8.80 & $0.044$ \\
\citetalias{stetson2019} $B-I$           & $-1.2$ & 13.0 & 9.00 & $-0.013$    & $-1.3$ & 12.5 & 9.10 & $0.017$ \\
\noalign{\smallskip}
   & \multicolumn{8}{c}{NGC\,362} \\
\citetalias{nardiello2018} $F606W-F814W$  & $-1.2$ & 10.5 & 9.00 & $0.010$    & $-1.3$ & 11.0 & 8.85 & $0.026$ \\
{\it Gaia} $G_\mathrm{BP}-G_\mathrm{RP}$  & $-1.2$ & 10.5 & 8.81 & $0.027$    & $-1.3$ & 10.5 & 8.80 & $0.059$ \\
\citetalias{stetson2019} $B-I$            & $-1.2$ & 10.5 & 9.00 & $0.006$    & $-1.3$ & 10.5 & 8.90 & $0.035$ \\
\noalign{\smallskip}
   & \multicolumn{8}{c}{NGC\,1261} \\
\citetalias{nardiello2018} $F606W-F814W$  & $-1.2$ & 10.0 & 16.60 & $0.000$   & $-1.3$ & 10.5 & 16.40 & $0.016$ \\
{\it Gaia} $G_\mathrm{BP}-G_\mathrm{RP}$  & $-1.4$ & 11.0 & 16.40 & $0.030$   & $-1.5$ & 10.5 & 16.70 & $0.063$ \\
\citetalias{stetson2019} $B-I$            & $-1.3$ & 10.5 & 16.70 & $-0.002$  & $-1.4$ & 10.5 & 16.30 & $0.029$ \\
\noalign{\smallskip}
   & \multicolumn{8}{c}{NGC\,5024} \\
\citetalias{nardiello2018} $F606W-F814W$  & $-1.9$ & 12.5 & 18.40 & $0.018$    & $-1.9$ & 12.5 & 18.30 & $0.028$ \\
{\it Gaia} $G_\mathrm{BP}-G_\mathrm{RP}$  & $-1.9$ & 13.0 & 18.20 & $0.030$    & $-1.9$ & 12.5 & 18.40 & $0.050$ \\
\citetalias{stetson2019} $B-I$            & $-1.9$ & 12.5 & 18.70 & $0.010$    & $-1.9$ & 13.0 & 18.30 & $0.017$ \\
\noalign{\smallskip}
   & \multicolumn{8}{c}{NGC\,5053} \\
\citetalias{nardiello2018} $F606W-F814W$  & $-2.3$ & 12.0 & 17.70 & $0.020$    & $-2.3$ & 12.5 & 17.60 & $0.025$ \\
{\it Gaia} $G_\mathrm{BP}-G_\mathrm{RP}$  & $-2.1$ & 12.5 & 17.10 & $0.027$    & $-2.0$ & 12.5 & 17.00 & $0.039$ \\
\citetalias{stetson2019} $B-I$            & $-2.3$ & 12.0 & 17.55 & $0.021$    & $-2.3$ & 13.0 & 17.20 & $0.022$ \\
\noalign{\smallskip}
   & \multicolumn{8}{c}{NGC\,5272} \\
\citetalias{nardiello2018} $F606W-F814W$  & $-1.5$ & 11.5 & 10.08 & $0.017$    & $-1.6$ & 11.5 & 10.05 & $0.032$ \\
{\it Gaia} $G_\mathrm{BP}-G_\mathrm{RP}$  & $-1.5$ & 11.5 & 10.00 & $0.022$    & $-1.6$ & 11.0 & 10.30 & $0.052$ \\
\citetalias{stetson2019} $B-I$            & $-1.6$ & 12.0 & 10.23 & $0.011$    & $-1.6$ & 11.5 & 10.10 & $0.029$ \\
\noalign{\smallskip}
   & \multicolumn{8}{c}{NGC\,5466} \\
\citetalias{nardiello2018} $F606W-F814W$  & $-2.0$ & 12.0 & 15.80 & $0.017$    & $-2.1$ & 12.0 & 16.15 & $0.029$ \\
{\it Gaia} $G_\mathrm{BP}-G_\mathrm{RP}$  & $-1.9$ & 12.0 & 15.40 & $0.031$    & $-1.9$ & 12.0 & 15.40 & $0.050$ \\
\citetalias{stetson2019} $B-I$            & $-2.1$ & 12.0 & 15.90 & $0.026$    & $-2.0$ & 12.0 & 15.80 & $0.029$ \\
\noalign{\smallskip}
   & \multicolumn{8}{c}{NGC\,5897} \\
\citetalias{nardiello2018} $F606W-F814W$  & $-1.9$ & 12.5 & 12.45 & $0.133$    & $-2.0$ & 12.5 & 12.75 & $0.143$ \\
{\it Gaia} $G_\mathrm{BP}-G_\mathrm{RP}$  & $-1.9$ & 13.0 & 12.60 & $0.137$    & $-1.9$ & 13.0 & 12.40 & $0.156$ \\
\citetalias{stetson2019} $B-I$            & $-1.9$ & 12.5 & 12.70 & $0.116$    & $-1.9$ & 13.0 & 12.50 & $0.125$ \\
\noalign{\smallskip}
   & \multicolumn{8}{c}{NGC\,5904} \\
\citetalias{nardiello2018} $F606W-F814W$  & $-1.2$ & 11.5 & 7.25 & $0.020$    & $-1.3$ & 12.0 & 7.05 & $0.040$ \\
{\it Gaia} $G_\mathrm{BP}-G_\mathrm{RP}$  & $-1.3$ & 11.5 & 7.27 & $0.042$    & $-1.4$ & 11.5 & 7.25 & $0.075$  \\
\citetalias{stetson2019} $B-I$            & $-1.3$ & 11.5 & 7.46 & $0.022$    & $-1.3$ & 11.5 & 7.25 & $0.040$ \\
\noalign{\smallskip}
   & \multicolumn{8}{c}{NGC\,6093} \\
\citetalias{nardiello2018} $F606W-F814W$  & $-1.7$ & 12.5 & 10.50 & $0.212$   & $-1.7$ & 12.5 & 10.30 & $0.224$ \\
{\it Gaia} $G_\mathrm{BP}-G_\mathrm{RP}$  & $-1.7$ & 12.5 & 10.55 & $0.210$   & $-1.7$ & 12.5 & 10.38 & $0.236$  \\
\citetalias{stetson2019} $B-I$            & $-1.7$ & 12.5 & 10.60 & $0.185$   & $-1.7$ & 12.5 & 10.40 & $0.203$ \\
\noalign{\smallskip}
   & \multicolumn{8}{c}{NGC\,6101} \\
\citetalias{nardiello2018} $F606W-F814W$  & $-1.9$ & 12.0 & 14.30 & $0.117$   & $-2.0$ & 12.5 & 14.35 & $0.126$ \\
{\it Gaia} $G_\mathrm{BP}-G_\mathrm{RP}$  & $-1.9$ & 12.0 & 14.00 & $0.135$   & $-2.1$ & 12.0 & 14.50 & $0.159$  \\
\citetalias{stetson2019} $B-I$            & $-2.0$ & 12.5 & 14.30 & $0.122$   & $-2.0$ & 13.0 & 13.80 & $0.134$ \\
\noalign{\smallskip}
   & \multicolumn{8}{c}{NGC\,6171} \\
\citetalias{nardiello2018} $F606W-F814W$  & $-0.9$ & 12.0 & 5.45 & $0.413$   & $-1.1$ & 12.0 & 5.47 & $0.436$ \\
{\it Gaia} $G_\mathrm{BP}-G_\mathrm{RP}$  & $-0.9$ & 12.0 & 5.45 & $0.415$   & $-1.1$ & 11.5 & 5.50 & $0.447$  \\
\citetalias{stetson2019} $B-I$            & $-1.0$ & 12.0 & 5.43 & $0.417$   & $-1.1$ & 11.5 & 5.44 & $0.445$ \\
\noalign{\smallskip}
   & \multicolumn{8}{c}{NGC\,6205} \\
\citetalias{nardiello2018} $F606W-F814W$  & $-1.5$ & 12.5 & 7.41 & $0.010$    & $-1.7$ & 13.0 & 7.38 & $0.034$ \\
{\it Gaia} $G_\mathrm{BP}-G_\mathrm{RP}$  & $-1.6$ & 12.5 & 7.40 & $0.025$    & $-1.6$ & 12.0 & 7.40 & $0.053$ \\
\citetalias{stetson2019} $B-I$            & $-1.6$ & 13.5 & 7.40 & $0.008$    & $-1.6$ & 13.0 & 7.39 & $0.025$ \\
\noalign{\smallskip}
   & \multicolumn{8}{c}{NGC\,6218} \\
\citetalias{nardiello2018} $F606W-F814W$  & $-1.2$ & 13.0 & 5.05 & $0.183$    & $-1.3$ & 13.0 & 5.06 & $0.202$ \\
{\it Gaia} $G_\mathrm{BP}-G_\mathrm{RP}$  & $-1.2$ & 13.5 & 4.80 & $0.197$    & $-1.3$ & 12.5 & 4.93 & $0.228$ \\
\citetalias{stetson2019} $B-I$            & $-1.2$ & 14.0 & 4.90 & $0.175$    & $-1.3$ & 12.5 & 5.03 & $0.212$ \\
\noalign{\smallskip}
   & \multicolumn{8}{c}{NGC\,6254} \\
\citetalias{nardiello2018} $F606W-F814W$  & $-1.5$ & 13.0 & 5.23 & $0.250$    & $-1.6$ & 13.0 & 5.17 & $0.270$ \\
{\it Gaia} $G_\mathrm{BP}-G_\mathrm{RP}$  & $-1.5$ & 13.0 & 5.05 & $0.257$    & $-1.6$ & 12.5 & 5.05 & $0.290$ \\
\citetalias{stetson2019} $B-I$            & $-1.7$ & 13.5 & 5.05 & $0.270$    & $-1.7$ & 13.0 & 5.12 & $0.285$ \\
\newpage
\noalign{\smallskip}
   & \multicolumn{8}{c}{NGC\,6341} \\
\citetalias{nardiello2018} $F606W-F814W$  & $-2.3$ & 12.5 & 8.41 & $0.030$    & $-2.3$ & 13.0 & 8.52 & $0.031$ \\
{\it Gaia} $G_\mathrm{BP}-G_\mathrm{RP}$  & $-2.2$ & 12.0 & 8.70 & $0.034$    & $-2.2$ & 12.5 & 8.53 & $0.051$ \\
\citetalias{stetson2019} $B-I$            & $-2.2$ & 13.0 & 8.45 & $0.022$    & $-2.3$ & 13.5 & 8.40 & $0.031$ \\
\noalign{\smallskip}
   & \multicolumn{8}{c}{NGC\,6352} \\
\citetalias{nardiello2018} $F606W-F814W$  & $-0.5$ & 12.5 & 5.37 & $0.228$    & $-0.7$ & 12.5 & 5.36 & $0.252$ \\
{\it Gaia} $G_\mathrm{BP}-G_\mathrm{RP}$  & $-0.6$ & 12.5 & 5.24 & $0.296$    & $-0.8$ & 13.0 & 5.20 & $0.321$ \\
\citetalias{stetson2019} $B-I$            & $-0.7$ & 12.0 & 5.34 & $0.314$    & $-0.8$ & 12.5 & 5.30 & $0.328$ \\
\noalign{\smallskip}
   & \multicolumn{8}{c}{NGC\,6362} \\
\citetalias{nardiello2018} $F606W-F814W$  & $-0.9$ & 12.0 & 7.77 & $0.050$    & $-1.1$ & 12.0 & 7.80 & $0.079$ \\
{\it Gaia} $G_\mathrm{BP}-G_\mathrm{RP}$  & $-0.9$ & 12.0 & 7.52 & $0.063$    & $-1.2$ & 12.0 & 7.75 & $0.115$ \\
\citetalias{stetson2019} $B-I$            & $-1.0$ & 12.0 & 7.73 & $0.054$    & $-1.1$ & 11.5 & 7.80 & $0.084$ \\
\noalign{\smallskip}
   & \multicolumn{8}{c}{NGC\,6366} \\
\citetalias{nardiello2018} $F606W-F814W$  & $-0.6$ & 11.5 & 3.37 & $0.718$    & $-0.7$ & 11.5 & 3.37 & $0.728$ \\
{\it Gaia} $G_\mathrm{BP}-G_\mathrm{RP}$  & $-0.5$ & 11.0 & 3.20 & $0.744$    & $-0.7$ & 11.0 & 3.34 & $0.760$ \\
\citetalias{stetson2019} $B-I$            & $-0.6$ & 11.5 & 3.28 & $0.762$    & $-0.7$ & 11.5 & 3.24 & $0.780$ \\
\noalign{\smallskip}
   & \multicolumn{8}{c}{NGC\,6397} \\
\citetalias{nardiello2018} $F606W-F814W$  & $-2.0$ & 13.0 & 2.50 & $0.182$    & $-2.0$ & 13.0 & 2.54 & $0.188$ \\
{\it Gaia} $G_\mathrm{BP}-G_\mathrm{RP}$  & $-1.8$ & 12.5 & 2.37 & $0.193$    & $-1.8$ & 12.0 & 2.44 & $0.213$ \\
\citetalias{stetson2019} $B-I$            & $-1.9$ & 13.5 & 2.42 & $0.184$    & $-2.0$ & 13.0 & 2.45 & $0.202$ \\
\noalign{\smallskip}
   & \multicolumn{8}{c}{NGC\,6541} \\
\citetalias{nardiello2018} $F606W-F814W$  & $-1.7$ & 13.0 & 7.56 & $0.114$    & $-1.8$ & 13.0 & 7.61 & $0.128$ \\
{\it Gaia} $G_\mathrm{BP}-G_\mathrm{RP}$  & $-1.8$ & 13.0 & 7.64 & $0.130$    & $-1.8$ & 13.0 & 7.55 & $0.153$ \\
\citetalias{stetson2019} $B-I$            & $-1.8$ & 13.5 & 7.50 & $0.123$    & $-1.9$ & 13.5 & 7.50 & $0.142$ \\
\noalign{\smallskip}
   & \multicolumn{8}{c}{NGC\,6723} \\
\citetalias{nardiello2018} $F606W-F814W$  & $-0.9$ & 12.0 & 8.21 & $0.052$    & $-1.1$ & 12.5 & 8.12 & $0.077$ \\
{\it Gaia} $G_\mathrm{BP}-G_\mathrm{RP}$  & $-1.1$ & 12.5 & 7.97 & $0.091$    & $-1.2$ & 12.5 & 8.07 & $0.116$ \\
\citetalias{stetson2019} $B-I$            & $-1.0$ & 12.5 & 8.07 & $0.057$    & $-1.1$ & 12.0 & 8.07 & $0.088$ \\
\noalign{\smallskip}
   & \multicolumn{8}{c}{NGC\,6752} \\
\citetalias{nardiello2018} $F606W-F814W$  & $-1.4$ & 13.0 & 4.11 & $0.048$    & $-1.6$ & 13.5 & 4.07 & $0.069$ \\
{\it Gaia} $G_\mathrm{BP}-G_\mathrm{RP}$  & $-1.4$ & 13.5 & 3.93 & $0.052$    & $-1.5$ & 12.5 & 4.03 & $0.086$ \\
\citetalias{stetson2019} $B-I$            & $-1.4$ & 14.0 & 3.98 & $0.034$    & $-1.6$ & 13.0 & 4.05 & $0.076$ \\
\noalign{\smallskip}
   & \multicolumn{8}{c}{NGC\,6779} \\
\citetalias{nardiello2018} $F606W-F814W$  & $-1.9$ & 12.0 & 10.70 & $0.245$    & $-2.0$ & 13.0 & 10.50 & $0.253$ \\
{\it Gaia} $G_\mathrm{BP}-G_\mathrm{RP}$  & $-2.1$ & 12.0 & 11.00 & $0.255$    & $-2.0$ & 12.5 & 10.70 & $0.268$ \\
\citetalias{stetson2019} $B-I$            & $-2.0$ & 12.0 & 10.80 & $0.246$    & $-2.0$ & 12.5 & 10.50 & $0.255$ \\
\noalign{\smallskip}
   & \multicolumn{8}{c}{NGC\,6809} \\
\citetalias{nardiello2018} $F606W-F814W$  & $-1.8$ & 12.0 & 5.36 & $0.116$    & $-1.9$ & 13.0 & 5.30 & $0.124$ \\
{\it Gaia} $G_\mathrm{BP}-G_\mathrm{RP}$  & $-1.8$ & 12.0 & 5.24 & $0.137$    & $-1.8$ & 12.0 & 5.27 & $0.156$ \\
\citetalias{stetson2019} $B-I$            & $-1.9$ & 12.5 & 5.38 & $0.123$    & $-1.8$ & 13.0 & 5.24 & $0.124$ \\
\noalign{\smallskip}
   & \multicolumn{8}{c}{NGC\,6838} \\
\citetalias{nardiello2018} $F606W-F814W$  & $-0.6$ & 12.0 & 4.04 & $0.202$    & $-0.8$ & 12.5 & 4.02 & $0.224$ \\
{\it Gaia} $G_\mathrm{BP}-G_\mathrm{RP}$  & $-0.6$ & 11.5 & 3.95 & $0.240$    & $-0.9$ & 12.5 & 4.00 & $0.269$ \\
\citetalias{stetson2019} $B-I$            & $-0.7$ & 12.0 & 4.00 & $0.244$    & $-0.8$ & 12.5 & 3.95 & $0.258$ \\
\noalign{\smallskip}
   & \multicolumn{8}{c}{NGC\,7099} \\
\citetalias{nardiello2018} $F606W-F814W$  & $-2.2$ & 12.5 & 8.38 & $0.043$    & $-2.2$ & 13.0 & 8.38 & $0.050$ \\
{\it Gaia} $G_\mathrm{BP}-G_\mathrm{RP}$  & $-2.0$ & 12.0 & 8.35 & $0.049$    & $-1.9$ & 12.0 & 8.24 & $0.065$ \\
\citetalias{stetson2019} $B-I$            & $-2.2$ & 13.0 & 8.36 & $0.045$    & $-2.1$ & 13.0 & 8.31 & $0.044$ \\
\hline
\end{longtable}

\section{Some more relations between the cluster parameters}
\label{more}

\begin{figure*}
   \centering
   \includegraphics[width=14.5cm, angle=0]{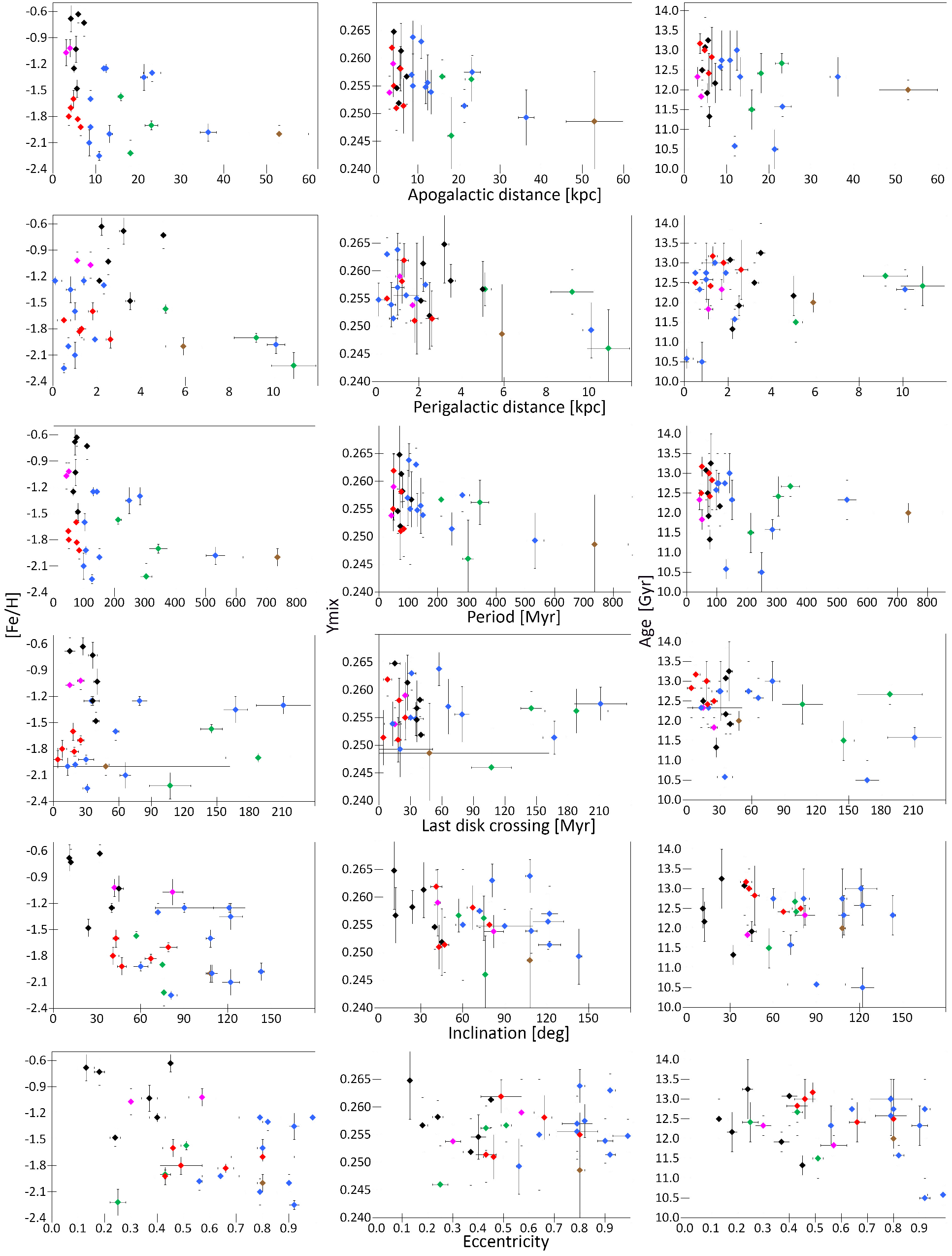}
   \caption{The relations between $[$Fe$/$H$]$ (left column), $Y_\mathrm{mix}$ (central column), and age (right column) vs the orbital parameters of the clusters.
   Clusters of different association are marked by different colours (see the text).}
\label{amr_orbit}
\end{figure*}

Fig.~\ref{amr_orbit} presents some more relations for the cluster parameters. A separation of the different groups of the clusters is evident in some diagrams.

\end{document}